\documentclass[journal,twoside,web]{ieeecolor}

\usepackage{booktabs}
\usepackage{subcaption}
\usepackage{multirow}
\usepackage{tikz}
\usepackage{textcomp}

\usepackage{jsen}
\usepackage{cite}
\usepackage{amsmath,amssymb,amsfonts,mathtools}
\usepackage{algorithmic}
\usepackage{graphicx}
\usepackage{textcomp}
\usepackage{wrapfig}

\renewcommand{\thetable}{\arabic{table}}

\DeclareMathOperator*{\argmin}{arg\,min}

\DeclareMathOperator*{\trace}{trace}
\DeclareMathOperator*{\blkdiag}{blkdiag}

\graphicspath{{./figures/}}
\newcommand{\eq}[1]{Eq.~\eqref{#1}}
\newcommand{\fig}[1]{Fig.~\ref{#1}}
\newcommand{\tab}[1]{Tab.~\ref{#1}}
\newcommand{\secref}[1]{Section~\ref{#1}}

\newcommand{\vv}[1]{\mathbf{#1}}
\newcommand{\smallsection}[1]{\noindent\textit{#1}.}

\newcommand{\mytexttilde}{{\raise.17ex\hbox{$\scriptstyle\mathtt{\sim}$}}}

\newcommand\copyrighttext{%
  \footnotesize This work has been submitted to the IEEE for possible publication. Copyright may be transferred without notice, after which this version may no longer be accessible.}
\newcommand\copyrightnotice{%
\begin{tikzpicture}[remember picture,overlay]
\node[anchor=south,yshift=10pt] at (current page.south) {\fbox{\parbox{\dimexpr\textwidth-\fboxsep-\fboxrule\relax}{\copyrighttext}}};
\end{tikzpicture}%
}

\def\BibTeX{{\rm B\kern-.05em{\sc i\kern-.025em b}\kern-.08em
    T\kern-.1667em\lower.7ex\hbox{E}\kern-.125emX}}
\definecolor{abstractbg}{rgb}{0.89804,0.94510,0.83137}
\begin{document}
\title{ORACLE: Occlusion-Resilient and Self-Calibrating mmWave Radar Network for People Tracking}

\author{Marco~Canil, \IEEEmembership{Graduate Student Member, IEEE}, Jacopo~Pegoraro, \IEEEmembership{Graduate Student Member, IEEE}, Anish~Shastri, \IEEEmembership{Graduate Student Member, IEEE}, Paolo~Casari \IEEEmembership{Senior Member, IEEE}, \\Michele~Rossi \IEEEmembership{Senior Member, IEEE}
\thanks{Marco Canil and Jacopo Pegoraro are with the Department of Information Engineering, University of Padova, Padova 35131, Italy (email: marco.canil@phd.unipd.it; jacopo.pegoraro@unipd.it).}
\thanks{Michele Rossi is with the Department of Information Engineering, University of Padova, Padova 35131, Italy, and with the Department of Mathematics ``Tullio Levi-Civita", University of Padova, Padova 35121, Italy (email: michele.rossi@unipd.it).}
\thanks{Anish Shastri and Paolo Casari are with the Department of Information Engineering and Computer Science, University of Trento, Povo 38123, Italy (email: anish.shastri@unitn.it; paolo.casari@unitn.it).}}

\IEEEtitleabstractindextext{%
\fcolorbox{abstractbg}{abstractbg}{%
\begin{minipage}{\textwidth}%
\begin{wrapfigure}[12]{r}{3in}%
\includegraphics[width=3in]{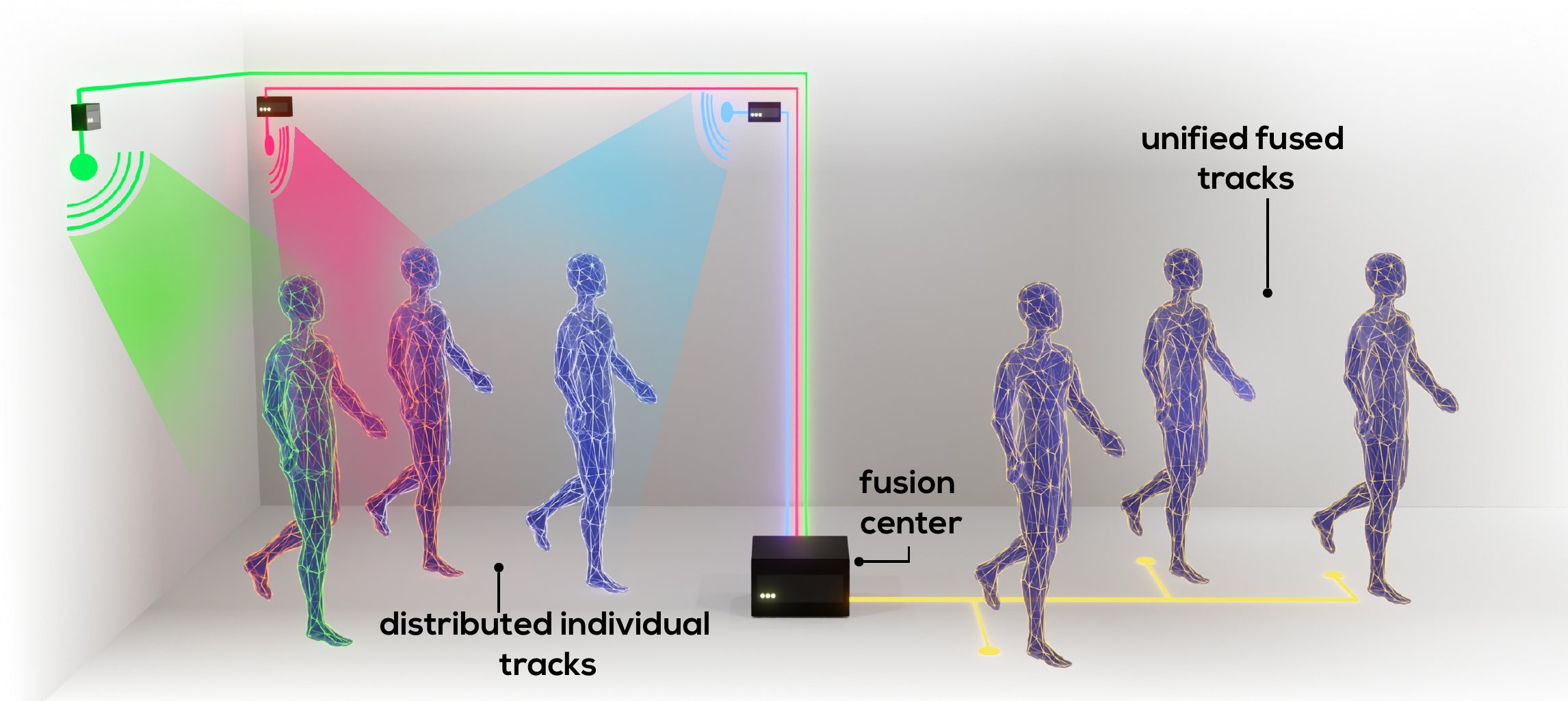}%
\end{wrapfigure}%
\begin{abstract}
Millimeter wave (mmWave) radar sensors are emerging as valid alternatives to cameras for the pervasive contactless monitoring of people in indoor spaces.
However, commercial mmWave radars feature a limited range (up to $6$--$8$~m) and are subject to occlusion, which may constitute a significant drawback in large, crowded rooms characterized by a challenging multipath environment.
Thus, covering large indoor spaces requires multiple radars with known relative position and orientation and algorithms to combine their outputs.
In this work, we present ORACLE, an autonomous system that \emph{(i)} integrates automatic relative position and orientation estimation from multiple radar devices by exploiting the trajectories of people moving freely in the radars' common fields of view, and \emph{(ii)} fuses the tracking information from multiple radars to obtain a unified tracking among all sensors.
Our implementation and experimental evaluation of ORACLE results in median errors of $0.12$~m and $0.03^\circ$ for radars location and orientation estimates, respectively. 
Fused tracking improves the mean target tracking accuracy by $27\%$, and the mean tracking error is $23$~cm in the most challenging case of $3$ moving targets.
Finally, ORACLE does not show significant performance reduction when the fusion rate is reduced to up to 1/5 of the frame rate of the single radar sensors, thus being amenable to a lightweight implementation on a resource-constrained fusion center.
\end{abstract}

\end{minipage}}}

\maketitle

\section{Introduction}
\label{sec:intro}
\copyrightnotice
\IEEEPARstart{R}{adars} operating in the mmWave frequency band have emerged as valid alternatives to cameras for indoor monitoring, as they are robust to changing and poor lighting conditions, and raise less privacy concerns~\cite{shah2019rf, savazzi2019use, shastri2022areview}. Their use enables advanced sensing applications spanning contactless people tracking \cite{pegoraro2021real}, personnel recognition \cite{zhao2019mid, pegoraro2021real, canil2022-milliTRACE-IR} and movement classification \cite{singh2019radhar}.
However, commercial mmWave radars have limited range~\cite{zhao2019mid} (up to \mbox{$6$--$8$~m}) and are subject to occlusion~\cite{pegoraro2021real}, which may constitute a significant drawback in large, crowded rooms containing furniture and walls. Thus, covering large indoor spaces requires multiple radars (i.e., radar networks), with \textit{known} relative position and orientation and algorithms to combine their output information.

In this work, we tackle the largely unexplored design of distributed mmWave radar networks to monitor people in indoor spaces. Our aim is to develop automatic calibration and sensor fusion algorithms to enable the quick deployment of multiple, jointly operating radars with \textit{no human intervention} and \textit{no accurate synchronization} between the devices.
Multistatic radars used in, e.g., \cite{shen2023indoor, bocus2021passive}, require synchronizing the devices' clocks in order to allow \textit{coherent} processing of the received signals. This is highly impractical, as it mandates clock distribution through an optical connection to be set up, jeopardizing the ease and speed of deployment. For this reason, we rather assume that the radar devices operate independently (i.e., they can only receive \textit{their own} transmitted signals), and communicate the result of the target detection and tracking steps to a fusion center.
In this scenario, we need to solve two main issues: \textit{(i)} automatically obtain the positions and orientations of the radars (\textit{self-calibration}), as they are often unknown, or it is impractical to measure them at deployment time; and \textit{(ii)} combine the environment perception capabilities of the multiple radars (\textit{sensor fusion}), so as to boost their sensing accuracy and mitigate occlusion. 
The few existing solutions to point \textit{(i)} present significant practicality and usability limitations in real scenarios \cite{li2021pedestrian, iwata2021multiradar, xie2022self}. Point \textit{(ii)}, instead, has not been investigated with indoor mmWave radars, to the best of our knowledge. This aspect is particularly challenging as we aim at enhancing the tracking accuracy \textit{without leveraging coherent processing}, thus only assuming coarse synchronization as provided by popular network protocols (e.g., the network time protocol, NTP).

In this paper we propose ORACLE, a solution to the mmWave radar network deployment and integration problem. Our contribution is twofold. As a first step, ORACLE automatically estimates the location and orientation of multiple radars with respect to a common coordinate system through an improved version of our previous work \cite{shastri2022mmSCALE}.
For this, ORACLE takes the trajectories of people moving in the environment as a reference. 
Then, the system fuses the information about moving people tracked by different radars at a fusion center (FC), enhancing the resilience of the tracking process in case of occlusion. 
ORACLE processes local information, transmitted by the radars to the FC, in a slotted-time fashion, thus handling the high variability in the frame rate of commercial radars. 
Then, it merges local tracks and provides a global representation of the moving targets in the environment.

The original contributions of this work are:
\begin{enumerate}
\item We propose ORACLE, a novel plug-and-play system for the real-time, automatic self-calibration and integration of multiple \textit{incoherent} mmWave radars for indoor people tracking.
\item As a first component of ORACLE, we present a fully-automated method for the self-calibration of multiple mmWave radars. The algorithm extends our previous work \cite{shastri2022mmSCALE} by adding a \emph{masking} phase (see \secref{sec:masking}) that handles a wider range of cases and provides better calibration results.
ORACLE estimates the relative positions and orientations of the radars with a median error of $0.12$~m and $0.03^\circ$, respectively, when 3 people move in the environment.
\item ORACLE includes a track-to-track radar fusion algorithm that combines information about the same subject collected by different radars. This improves the mean tracking accuracy by up to $27\%$ with respect to single-sensor tracking. %
\item We evaluate our method via an extensive measurement campaign through the RadNet platform \cite{bashirov2022RadNet}, using $4$ commercial mmWave radars deployed in realistic conditions and multiple subjects, including challenging human motion. In the most difficult case of $3$ subjects moving concurrently, ORACLE achieves a tracking accuracy of $87\%$ and a mean tracking error of $23$~cm.

\end{enumerate}
The remainder of this paper is organized as follows: \secref{sec:rel-work} provides a summary of the related work. In \secref{sec:problem}, the challenges of designing a mmWave radar network are introduced. \secref{sec:approach} presents and discusses ORACLE, the proposed method. \secref{sec:results} provides some insights regarding the practical implementation of ORACLE and presents the experimental results on our testbed. Finally, concluding remarks are drawn in \secref{sec:concl}.

\section{Related work}
\label{sec:rel-work}
\smallsection{Multistatic radars} Using multiple radar receivers with synchronized clocks enables coherent analysis of the received signals, which yields significant processing gains due to spatial diversity \cite{chernyak1998fundamentals}. Existing works have leveraged this principle for drone detection \cite{hoffmann2016micro} and people tracking \cite{ledergerber2020multi,bartoletti2014sensor,he2012range, shen2023indoor}. Despite their potentially superior accuracy and resolution, the main drawback of multistatic radars lies in their impracticality and deployment cost. Indeed, a common clock source needs to be distributed to the receivers, either via optical links or GPS, which is not available indoors. This would prevent the radar sensors from being quickly deployed, used, or relocated. Conversely, we target a scenario where ease of deployment and minimal human intervention are key requirements. For this reason, multistatic radars are not applicable and ORACLE focuses on track-level sensor fusion from incoherent sensors, that are only coarsely synchronized (i.e., not at the clock level) using standard NTP.

\smallsection{Radar networks}

A large body of work has considered the use of incoherent radar networks in airborne and automotive applications to improve the detection and tracking capabilities of the standalone sensors, e.g., 
\cite{shu2007data, folster2005data}. These works tackle the fusion of distributed radar tracks without leveraging the multistatic gain available with precise synchronization.
However, the considered radar setups significantly differ from mmWave radar network deployments which typically take place indoors or in short-range ($6$--$8$~m) outdoor scenarios, where the presence of multiple subjects may create crowded scenes and occlusions. The latter occur when people or objects block the line of sight between a radar and the target.
Moreover, most existing works on mmWave radar networks rely on very simple offline data association rules based on \textit{known} sensor positions, with no data fusion to improve the tracking accuracy \cite{guendel2021continuous, meng2020gait}. To the best of our knowledge, only one work \cite{xie2022self} has addressed indoor radar networks for people tracking, although using a lower frequency band ($7$--$10$ GHz). A major drawback of \cite{xie2022self} is the assumption that \textit{only one person} is present in the environment, which is unrealistic in general indoor settings.
All the above limitations are solved by the proposed solution, which is the first system that \textit{(i)} combines the information from \textit{multiple radars}, handling the presence of \textit{multiple subjects}, \textit{(ii)} automatically estimates their relative positions and orientations, and \textit{(iii)} shows robust real-time performance thanks to its low complexity and distributed computation load.

\smallsection{Radar networks self-calibration} 
To our knowledge, only two works have tackled the problem of self-calibration in mmWave radar networks, i.e.,~\cite{li2021pedestrian, iwata2021multiradar}. Both have significant practical limitations: \cite{li2021pedestrian} requires that just a single subject, following a linear walking trajectory, appears in the field of view (FoV) of the radars, while \cite{iwata2021multiradar} can handle multiple subjects, but all of them need to be static (e.g., sitting). Such assumptions considerably limit the application scope of these systems. Conversely, our method completely automates the calibration process, working with movement trajectories of arbitrary shapes and with multiple concurrently moving targets. Actually, ORACLE \textit{benefits} from having multiple trajectories of complex and irregular shape that span a large portion of the FoVs of the radars, as they lead to a more accurate calibration.

\section{Problem outline}
\label{sec:problem}

In this section, we first present an overview of mmWave Multiple-Input Multiple Output (MIMO) radars. Then, we formalize the problems of combining the information obtained by the different radars in the network at a central fusion entity, and of estimating their relative positions and orientations.

\subsection{mmWave MIMO radars}
\label{sec:prel}

A MIMO FMCW radar jointly estimates the distance, the radial velocity, and the angular position of the targets with respect to the radar itself~\cite{patole2017automotive}. During the sensing process, the radar transmits sequences of linear chirp signals with bandwidth $\mathcal{B}$. A full sequence, or ``radar frame", is repeated with a period of $T_s$ seconds.
The distance, $r$, and velocity, $v$, of the targets are computed from the frequency shift induced by the delay of each reflection, usually by applying discrete Fourier transform (DFT) processing. The FMCW radar distance resolution is related to the bandwidth $\mathcal{B}$ by $\Delta r = c/(2\mathcal{B})$, where $c$ is the speed of light. This makes \mbox{mmWave} devices accurate to the level of a few centimeters using a bandwidth of $2$--$4$~GHz~\cite{pegoraro2021real}.
Furthermore, using a 2D array of multiple receiving antennas makes it possible to obtain the angle of arrival (AoA) of the reflections along the azimuth ($\theta$), and the elevation ($\phi$) domains, by leveraging phase shifts across different antenna elements. The azimuthal AoA resolution depends on the number of antennas $N$ in the array and is given by $\Delta \theta = \lambda/(Nd\cos\theta)$, where $d$ is the spacing between the antennas.
Due to the high ranging resolution, a human presence in the environment generates a large number of reflecting points, which are detected by the radar. 
This set of points, usually termed {\it radar point cloud}, can be transformed into the $3$-dimensional Cartesian space using the distance, azimuth, and elevation angles information of the multiple body parts. 
Each point is described by a vector \mbox{$\left[x, y, z\right]^T$} including the point's spatial coordinates $x,y,z$ obtained by transforming $r, \theta$ and $\phi$.
Movement trajectories can be tracked across time from the point-clouds. 

\subsection{Sensor fusion in mmWave radar networks}
\label{sec:fusion-problem}

Consider a mmWave radar network consisting of $S$ monostatic radar sensors. Each radar has local computational capabilities and a communication interface that enables them to transmit information to a FC. The sensors are identified by indices $s=1, \dots, S$, while quantities related to the FC are denoted by superscript $^c$. All radar sensors operate at discrete time steps of duration $T_s$, indexed by variable $k$. 
The FC also operates at discrete time steps that, in general, may have a different duration $T_c$ and are indexed by variable $m$.

The people tracking problem relates to estimating the subjects' movement trajectories in the $(x, y)$ horizontal plane across time, exploiting the measurements of the multiple radar sensors.
For this, we define the \textit{state} of subject $u$, seen by the FC at time $m$, as \mbox{$\vv{x}_m(u) = \left[ x_m(u), y_m(u), \dot{x}_m(u), \dot{y}_m(u) \right]^T$}, containing $u$'s coordinates and the corresponding velocity components $\dot{x}_m(u)$ and $\dot{y}_m(u)$. 
We assume that the state's evolution obeys a constant-velocity (CV) model~\cite{wagner2017radar}.
At the FC, the state model for target $u$ is
\begin{equation}\label{eq:state-evol}
    \vv{x}_m(u) = \vv{F}_{T_c}\vv{x}_{m-1}(u) + \vv{w}_m(u),
\end{equation}
where $\vv{F}_{T_c}$ is the state transition matrix that projects the state forward by a time duration $T_c$, according to the CV model, while $\vv{w}_m(u)$ is the (global) Gaussian process noise, having zero-mean and covariance matrix $\vv{W}$ \cite{pegoraro2021real,chong2000architectures}. The process noise is here considered to be generated by a random acceleration that is not explicitly accounted for by the CV model \cite{pegoraro2021real}. 
Sensor measurements of the state of target $n$, at time $k$, are obtained according to
\begin{equation}\label{eq:state-obs}
    \vv{z}^s_k(n) = \vv{H}\vv{x}_{k}(n) + \vv{v}^s_k(n), \quad s = 1, \dots, S,
\end{equation}
where $\vv{z}^s_k(n)$ is the observation obtained from sensor $s$, $\vv{H}$ is the observation matrix relating the observation to the state and $\vv{v}^s_k(n)$ is the sensor-specific measurement noise having covariance matrix $\vv{V}_k$ \cite{pegoraro2021real}. In our system, all sensors are of the same type and have the same specifications. Therefore, we can safely assume that the measurement error processes have the same zero-mean, Gaussian distribution, whose covariance is time-varying due to the dependence of the radars' resolution on the position of the targets in the FoV \cite{pegoraro2021real}.

The aim of our system is to estimate $\vv{x}_m$ over time, exploiting the measurements collected by the sensors. Note that: 
\textit{(i)} the correspondence between the targets tracked by each sensor is \textit{unknown}, and finding a suitable association between them is part of the problem we tackle;
\textit{(ii)} the algorithm can handle the fact that the same target may be tracked simultaneously by one or more sensors; and \textit{(iii)} the sensors may collect the measurements and obtain state estimates at a different time granularity than that of the FC.

Each sensor locally tracks the targets in the environment. The common approach to people tracking from mmWave radar point-clouds~\cite{zhao2019mid, meng2020gait, pegoraro2021real} includes: \textit{(i)} a detection phase via density-based clustering algorithms (e.g., DBSCAN~{\cite{ester1996density}}) to separate the reflections from multiple subjects; \textit{(ii)} applying Kalman filtering techniques \cite{kalman1960new} on each cluster centroid to track the movement trajectory of each subject in space.
The KF used at each sensor provides, at each time step, an estimate of the state of the targets in its FoV and the corresponding error covariance. We call $N_k^s$ the number of such targets at time $k$, $\hat{\vv{x}}_{k|k}^s(n)$ the estimated state of target $n$ after the KF update step, and $\vv{C}_{k|k}^s(n)$ the associated error covariance. Note that the above quantities are sensor-dependent, as different sensors provide their own estimates of the state of the same target. 
We denote by $\mathcal{T}^s_k(n) = \{\hat{\vv{x}}_{k|k}^s(n), \vv{C}_{k|k}^s(n)\}$ the track corresponding to target $n$ as estimated by sensor $s$, expressed with respect to \textit{its own} reference frame. In addition, we assume that each sensor is able to provide a timestamp, $\tau_k^s$, corresponding to the current time step, according to its local time reference (e.g., internal clock or network time). For the timestamps to match between different sensors, some level of synchronization is needed within the radar network (e.g., using NTP).  At the end of each time step, sensor $s$ transmits the set of its tracks, denoted by $\boldsymbol{\mathcal{T}}^s_k = \{\mathcal{T}^s_k(1), \dots, \mathcal{T}^s_k(N_k^s), \tau^s_k\}$ to the FC, together with the corresponding timestamp. 

If the same target is tracked by more than one sensor, the FC should maintain a single track for it, which is updated and improved by fusing the information coming from the sensors.
Our aim is to develop an algorithm to estimate the position of the targets across time at the FC, in the form of \textit{central tracks} $\mathcal{T}^c_m(u) =  \{\hat{\vv{x}}_{m|m}^c(u), \vv{C}_{m|m}^c(u)\}$, obtained by combining the sensor information $\boldsymbol{\mathcal{T}}^s_k, \, s=1, \dots, S$. The above problem is complicated by \textit{correlation} between the estimation errors of the tracks obtained at the sensors and at the FC. 
From \eq{eq:state-evol}, one can see that some correlation exists between all the tracks that refer to the same target (also in different sensors), as the process noise is the same, but this can be typically neglected if the process noise has low intensity or if the radar measurement rate is high with respect to the subject's motion \cite{chong2000architectures}.
Conversely, the error correlation between a central track and a sensor track of the same target cannot be ignored, as the FC obtains its own tracks as a function of the sensor tracks.
This is especially true for our real-time application, where the fusion occurs frequently, e.g., from $10$ to $15$ times per second.

\begin{figure}
	\begin{center}   
		\includegraphics[width=8.4cm]{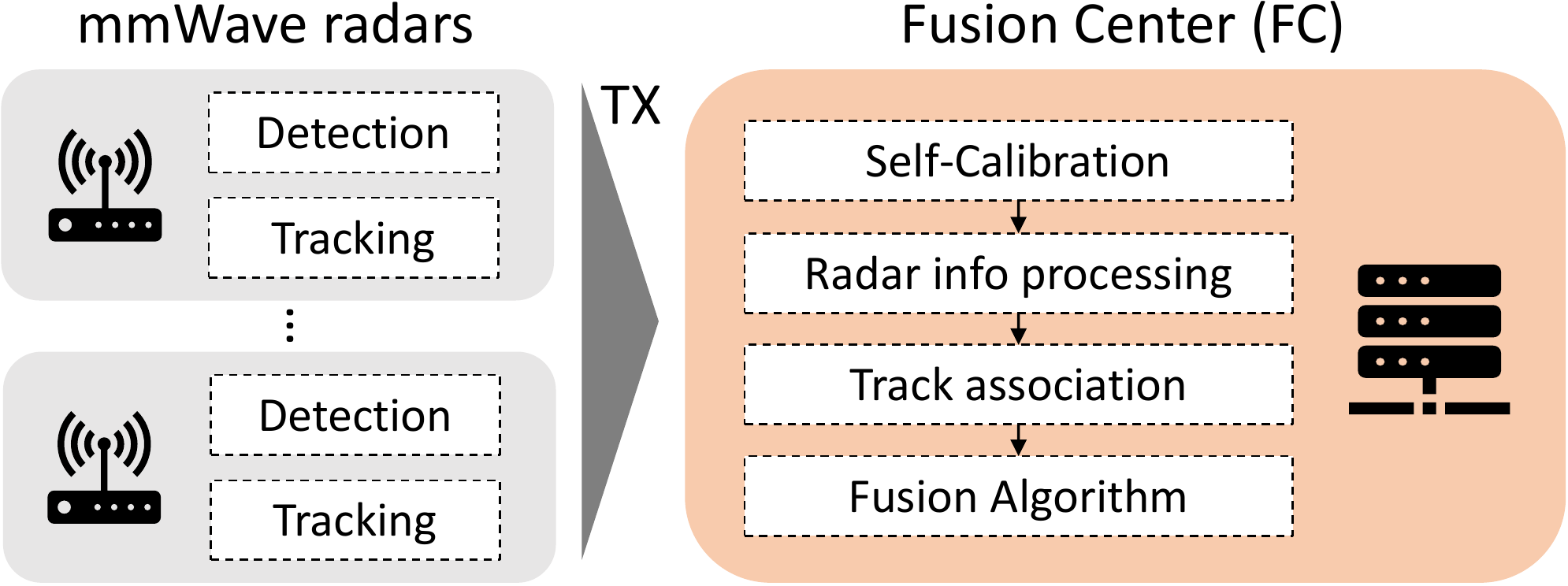} 
		\caption{Proposed workflow.
		}
		\label{fig:workflow}
	\end{center}
\end{figure}

\subsection{Self-calibration of mmWave radar networks}
\label{sec:selfcal-problem}
The track sets that the radars transmit to the FC are expressed in the local reference frames of the sensors. Any algorithm that fuses them to improve the tracking accuracy requires to know the sensors' relative position and orientation. However, manually measuring them is impractical and prone to errors, therefore an automatic self-calibration procedure is highly appealing.
Here, we propose to exploit the trajectories of \emph{targets of opportunity} that move within the radars' FoVs, independently tracked by each sensor.
Tracks from different radars that correspond to the same target have almost the same \emph{shape}, up to a rigid transformation and some noise.
Estimating such rigid transformation parameters corresponds to estimating the sensors' relative position and orientation.

Considering the system of $S$ radars, deployed in the same area, call \mbox{$\mathcal{F}_s, \, s=1, \dots, S$}, their reference systems (RSs). Each RS consists of a pair \mbox{$\mathcal{F}_s = \{\vv{t}_s,\vv{R}_s\}$}, where \mbox{$\vv{t}_s$} is the $2 \times 1$ vector with the coordinates of the $s$-th RS's origin and $\vv{R}_s$ is the \mbox{$2 \times 2$} rotation matrix specifying its orientation.
Without loss of generality, in this paper we consider a \textit{global} RS (of the FC) which coincides with that of radar $1$ and for which it holds that \mbox{$\vv{t}_1 = \boldsymbol{0}_{2 \times 1}$} and \mbox{$\vv{R}_1 = \vv{I}_2$}, respectively, the $2 \times 1$ zero vector and the rank $2$ identity matrix. 
Self-calibrating the system consists in estimating $\mathcal{F}_s, \, s=2, \dots, S$.
We define the \textit{movement trajectory} of target $n$, as seen by sensor $s$, as the sequence of position estimates of the target, $\hat{\vv{p}}^s_k(n) = [\hat{x}^s_k(n), \hat{y}^s_k(n)]^T$, for $k = 1, \dots, \, K$, where $k$ is the discrete time index. 
Note that $\hat{\vv{p}}^s_k(n)$ contains the first two components of the KF state estimate $\hat{\vv{x}}^s_{k|k}(n)$. 
An estimate of the rotation matrix and of the translation vector between radar $s$ and radar $1$ (our reference) can be obtained solving the following Least-Squares (LS) problem

\begin{equation}
\label{eq:opt-problem}
 \argmin_{\substack{\vv{R}_s \in SO(2) \\ \vv{t}_s \in \mathbb{R}^2}}
    \sum_{k=1}^{K} \left|\left|(\vv{R}_s \hat{\vv{p}}^s_{k}(n) + \vv{t}_s) - \hat{\vv{p}}^1_{k}(n)\right|\right|_2, 
\end{equation}
where $SO(2)$ denotes the special orthogonal group in dimension $2$ (i.e., the set of all possible rotations around a point in a 2-dimensional space) and $||\cdot||_2$ is the Euclidean norm.
While the translation vector of sensor $s$ with respect to the global RS is directly obtained by solving \eq{eq:opt-problem}, the orientation angle, denoted by $\theta_s$, is given by $\theta_s = \cos^{-1}\left({ \trace(\vv{R}_s)}/2\right)$.

In \secref{sec:selfcal-method} we present the proposed approach to solve the self-calibration problem in the more complex and realistic scenario where: \textit{(i)} multiple sensors concurrently track multiple targets;
\textit{(ii)} the track-target correspondence among different sensors is \textit{unknown}, so, an association strategy has to be developed; and 
\textit{(iii)} the trajectories should be aligned in time before using \eq{eq:opt-problem}.

\section{Proposed approach}
\label{sec:approach}
In this section, we first present a high-level overview of the processing blocks of ORACLE and then provide a detailed description of each of them. \fig{fig:workflow} presents the workflow of ORACLE.

\smallsection{Self-calibration} In this phase, the relative positions and orientations of the radars are obtained from the trajectories of targets of opportunity (see \secref{sec:selfcal-method}). The steps are:
\begin{enumerate}
    \item \textit{Time alignment.}
    A time alignment between the trajectories from radar $1$ and the trajectories from radar $s$ is sought, by minimizing the distance between the trajectories' timestamps (see \secref{sec:time-alignment}).
    \item \textit{Track association.} Using the time alignment from point $1$), we solve the problem in \eq{eq:opt-problem} for all the trajectory pairs and compute a corresponding association cost matrix. Using the cost matrix, the best associations between track pairs are computed (see \secref{sec:tracks-association}).   
    \item \textit{Masking.} When estimating the roto-translation parameters at point $4$), multiple track pairs will be used all together. 
    To avoid that possibly wrong associations spoil the final results, all possible subsets of the best associations from point $2$) are considered through a \emph{masking} operation and a new association cost is computed using all the trajectories in each subset. In the end, the subset giving the lowest cost will be selected for the final parameters estimation in point $4$) (see \secref{sec:masking}).  
    \item \textit{Radar calibration.} All the track pairs from point $3$) are stacked together and used to set up a rigid transformation problem as in \eq{eq:opt-problem} that provides the final position and orientation estimates for the radar (see \secref{sec:calibration}).

\end{enumerate}
\smallsection{Multi-radar fusion} Here, the tracks from the radars are fused at the FC to build a set of \textit{central} tracks associated with the subjects in the environment (see \secref{sec:multirad-fusion}). This includes:
\begin{enumerate}
    \item \textit{Slotted sensor information processing.} The tracks information from the sensors are sent to the FC and processed using a slotted protocol. (see \secref{sec:slotted-proc}).
    \item \textit{Track association.} A method to associate (frame-by-frame) those tracks that correspond to the same target, according to their statistical similarity, is used to select pairs of tracks to be fused (see \secref{sec:tt-assoc}).
    \item \textit{Radar track fusion algorithm.} 
    The fusion algorithm combines sensor tracks with the central tracks using different rules depending on the type of fusion event (see \secref{sec:track-fusion}).
\end{enumerate}

\subsection{Self-calibration}\label{sec:selfcal-method}
\subsubsection{Time alignment}\label{sec:time-alignment}
For simplicity of notation, call $\vv{n}_1$ a trajectory from sensor $1$ and $\vv{n}_s$ a trajectory from sensor $s$. Each of them contains the sequence of the position estimates of some target that may, or may not, coincide.
Sensors communicate position estimates to the FC along with their timestamps.
Note that the trajectories may have a different length.
The time alignment is then performed so that each position estimate of trajectory $\vv{n}_1$ is associated with the position estimate of trajectory $\vv{n}_s$ that minimizes the time difference between the two acquisition instants.
Elements of trajectory $\vv{n}_1$ that do not have a corresponding element of trajectory $\vv{n}_s$ within $T_c$ seconds are discarded and vice-versa (recall that $T_c$ is the duration of a FC time step).
This operation reduces the trajectories to a common length of $K$ time-aligned positions.
Call $\vv{k}_1$ and $\vv{k}_s$ the vectors containing the indices that provide the time-aligned sequences from radars $1$ and $s$, respectively. Note that $\vv{k}_1$ and $\vv{k}_s$ have the same length.
With the time alignment operation, we retain only the portions of the trajectories that are sufficiently well synchronized, in order to avoid performing the rigid transformation on wrongly associated points.
Once the trajectory association has been established, we define the \emph{mean time shift} of the pair \mbox{$\{\vv{n}_1, \vv{n}_s\}$} as
$\bar{\tau}{(\vv{n}_1, \vv{n}_s)} = \frac{1}{K} \sum_{k=1}^{K} |\tau_{\vv{k}_{1,k}}^{1} - \tau_{\vv{k}_{s,k}}^{s}|$, where $\vv{k}_{j,k}$ denotes the $k$-th element of vector $\vv{k}_j$.
The value of $\bar{\tau}{(\vv{n}_1, \vv{n}_s)}$, expressed in seconds, is related to the alignment quality of the two trajectories and will be used in the association step (see \secref{sec:tracks-association}).

\subsubsection{Track association}
\label{sec:tracks-association}

Our data association strategy consists in computing a \textit{cost} for each pair \mbox{$\{\vv{n}_1, \vv{n}_s\}$} and solving the resulting combinatorial cost minimization problem to obtain the best associations.
We assume to have $N_1$ and $N_s$ trajectories available at radars $1$ and $s$, respectively. Our cost function incorporates different aspects: \emph{(i)} the length of the trajectories, as longer trajectories are assumed to provide a better calibration; \emph{(ii)} the time alignment of the trajectories, as we should compare position estimates acquired almost simultaneously by the different radars; and \emph{(iii)} the quality of the rigid transformation, in terms of residual error in superimposing trajectories from the different radars.
We define the association cost, $A$, for the pair \mbox{$\{\vv{n}_1, \vv{n}_s\}$}, as
\begin{equation}
\label{eqn:cost_function}
    A(\vv{n}_1, \vv{n}_s) = -\rho(K, \bar{\tau}) \left[1 + \xi(\vv{n}_1, \vv{n}_s)\right]^{-1},
\end{equation}
where $\xi(\vv{n}_1, \vv{n}_s)$ is the sum of the LS residuals, after applying the time alignment and the rigid transformation, while $\rho(K, \bar{\tau})$ is a factor that favors trajectory pairs with a long overlap and a low mean time shift. 
The rigid transformation parameters $\vv{R}^{(\vv{n}_1, \vv{n}_s)}_s, \vv{t}^{(\vv{n}_1, \vv{n}_s)}_s$ are computed, using the time-aligned track pairs from point $1$), solving the LS problem in \eq{eq:opt-problem} in closed-form through a Singular-Value Decomposition (SVD) method \cite{sorkine2017least}. 
Then, the LS residuals sum is computed as $\xi(\vv{n}_1, \vv{n}_s) = \sum_{k=1}^{K} ||\,\vv{n}_{1, k} - \vv{R}^{(\vv{n}_1, \vv{n}_s)}_s\vv{n}_{s, k} - \vv{t}^{(\vv{n}_1, \vv{n}_s)}_s\,||_2$, where $k$ indexes trajectory positions.
Recalling that $T_c$ is the sampling interval of the FC, the corrective term is formalized as
$\rho(K, \bar{\tau}) = \ln(KT_c) \left[1 + \bar{\tau}(\vv{n}_1, \vv{n}_s)\right]^{-1}$.
Costs $A(\vv{n}_1,\vv{n}_s)$
are arranged into a \mbox{$N_1 \times N_s$} cost matrix
 and the optimal association of trajectories is obtained by minimizing the overall cost, computed through the Hungarian algorithm~\cite{kuhn1955hungarian}. This yields $N_t = \min \left(N_1, N_s\right)$ pairs of associated trajectories, which are possibly the same targets seen by the two radars. Due to the presence of spurious trajectories, ghost targets and clutter, we select a subset of the associated trajectory pairs that have a cost below a threshold $A_{\rm self}$, which represents a confidence value under which the pair is truly a trajectory pair generated by a human. Denote the set of selected track pairs by $Q^{1s}=\{\{\vv{s}_1^1, \vv{s}_1^s\}, \dots, \{\vv{s}_{N_t}^1, \vv{s}_{N_t}^s\}\}$, where $\vv{s}_p^q$ indicates the $p$-th selected track from radar $q$. In our experiments, we empirically adopted $A_{\rm self}=18$.

\subsubsection{Masking}
\label{sec:masking}
In phase 4) (see \secref{sec:calibration}), one or more of the $N_t$ track pairs selected during the previous phase are used combinedly to compute the final self-calibration parameters. Ideally, each of the selected track pairs should provide two trajectories, one from each radar, corresponding to the same target. However, in practice, there might be wrong associations. To mitigate this shortcoming, in phase 3) all possible subsets of the selected $N_t$ tracks are considered, through a \emph{masking} operation on the one-to-one track associations.
We call it a masking operation as it corresponds to purposedly ignoring (\textit{masking}) some of the track associations in the computation of the self-calibration parameters.
The same association cost of \eq{eqn:cost_function} is computed by stacking together all the trajectories in each subset. Then, the subset providing the lowest cost is used for the final calibration.
Formally, let $P(Q^{1s})$ be the set of all possible subsets 
of $Q^{1s}$ excluding the empty set. 
Recall that $Q^{1s}$ is the set of all selected track pairs after the track association phase.
Each element of $P(Q^{1s})$ is a set of track pairs from radar $1$ and radar $s$. 
For each element of $P(Q^{1s})$, all the trajectories from sensor $1$ are stacked in vector $\vv{q}_1$ and all the trajectories from sensor $s$ are stacked in vector $\vv{q}_s$ and the same operation is performed with the corresponding timestamp sequences. Then, cost $A(\vv{q}_1, \vv{q}_s)$ is computed as in \eq{eqn:cost_function} and all costs are stored in a matrix of dimension $(2^{N_t}-1) \times 1$.
The element of $P(Q^{1s})$ providing the lowest cost is selected. The $N_t^* \le N_t$ trajectory pairs contained in such minimum-cost element are used in phase 4) to compute the self-calibration parameters.
Since the masking phase cost is exponential in the number of track-to-track associations, it is possible to limit the maximum number of track pairs to be retained from $Q^{1s}$.
In this case, the track pairs with the highest cost are to be excluded.
In our experiments, we used a maximum of $5$ track pairs.

\subsubsection{Radar calibration}
\label{sec:calibration}
The $N_t^*$ trajectory pairs selected during the masking phase are then stacked together and used to set up a rigid transformation problem as in \eq{eq:opt-problem}.
The problem is solved with the same procedure described in \secref{sec:tracks-association} \cite{sorkine2017least}, obtaining the final rotation matrix and translation vector to calibrate radar $s$, namely $\{\vv{R}_s^*, \vv{t}_s^*\}$.
This step exploits all the available information from multiple subjects, improving the calibration accuracy by increasing the number of useful measurements per time frame.
Note that, even though target occlusion events may split a trajectory into multiple components, our algorithm still works by exploiting each resulting sub-trajectory.

\subsection{Multi-radar fusion}
\label{sec:multirad-fusion}

\begin{figure}
	\begin{center}   
		\includegraphics[width=8.4cm]{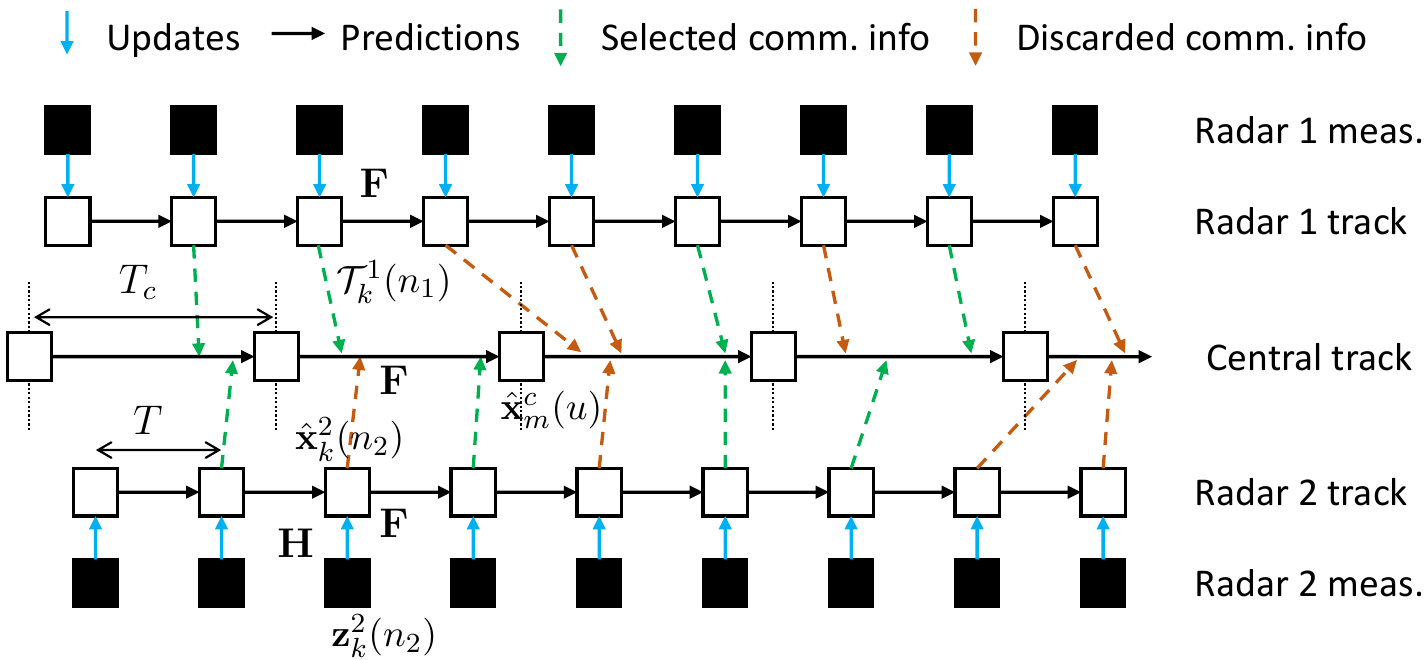} 
		\caption{Scheme of the proposed fusion algorithm with $2$ radars.
		}
		\label{fig:fusion}
	\end{center}
\end{figure}

Once the radar network is calibrated, i.e., we have an estimate $\{\vv{R}_s^*, \vv{t}_s^*\}, \forall s$, we can fuse the information coming from the $S$ radars at the FC.
In the following, we consider $S=2$ for better clarity in the algorithm description, but the method works for an arbitrary $S$. 
We denote the \textit{precision matrix}, which is defined as the inverse of a covariance matrix, by $\vv{P} = \vv{C}^{-1}$.
The fusion algorithm, represented in \fig{fig:fusion}, is described as follows:

\subsubsection{Slotted sensor information processing}\label{sec:slotted-proc} The FC maintains a central time variable, denoted by $\tau_m^c = \tau_0 + mT_c$, which is incremented at the end of each central time step and where $\tau_0$ is the time when the FC starts operating.
In order to cope with the random variations in the sensor acquisition, processing, and communication times,
the FC operates on time slots of duration $T_c$. Specifically, several track sets from different sensors can be received during time step $m$ due to differences between the FC and the sensors' time steps and the variable communication time. Using the timestamp information contained in the received tracks, at time $m$ the FC filters out all the track sets that are not received within the interval $(\tau_{m-1}^c, \tau_m^c]$ and retains only the most recent track set from each sensor.
Formally, for each $s$, we select the track set whose timestamp is the solution to $\argmin_{\tau_{k}^s}(\lvert\tau_{m}^c - \tau_{k}^s\rvert)$. 
In the following, to highlight that a track set has been selected from sensor $s$ to be processed in time step $m$, we denote it as $\boldsymbol{\mathcal{T}}^s_m$, using the time index of the FC rather than that of the sensor, and we do the same for all the tracks it contains.

The slotted processing procedure \textit{(i)} reduces the number of fusion steps the FC carries out, using only the most recent information available from each sensor, and \textit{(ii)} avoids erroneously fusing outdated tracks.%

After selecting the sensor tracks, the FC transforms them to match its own reference system, using the information about the location and orientation of the radar sensors.
Then, according to the CV model, the tracks are propagated to the FC current time.
We denote by $\bar{\mathbf{t}}^*_s = [\mathbf{t}^*_s, \boldsymbol{0}_{1\times 2}]^T$ the augmented translation vector and by $\bar{\mathbf{R}}_s^* = \blkdiag(\mathbf{R}_s^*, \mathbf{R}_s^*)$
the $4 \times 4$ augmented rotation matrix of sensor $s$, where $\blkdiag(\cdot)$ returns a block diagonal matrix of its inputs. The transformation and propagation are performed, together, as
\begin{equation}\label{eq:sens-conv-state}
    \hat{\mathbf{x}}_m^s(n) =  \vv{F}_{\tau_m^c - \tau_k^s}\left[\bar{\mathbf{R}}_s^*\hat{\mathbf{x}}_m^{s'}(n) + \bar{\mathbf{t}}^*_s\right],
\end{equation}
\begin{equation}\label{eq:sens-conv-cov}
    \mathbf{C}_m^s(n)  = \vv{F}_{\tau_m^c - \tau_k^s}\bar{\mathbf{R}}_s^*\mathbf{C}_m^{s'}(n) \left(\bar{\mathbf{R}}_s^*\right)^T\vv{F}_{\tau_m^c - \tau_k^s}^T, 
\end{equation}
with $\hat{\mathbf{x}}_m^{s'}(n)$ and $\mathbf{C}_m^{s'}(n)$ being the state and covariance communicated by sensor $s$ that have been selected in the current central slot, expressed in the reference frame of sensor $s$, while $\hat{\mathbf{x}}_m^{s}(n)$ and $\mathbf{C}_m^{s}(n)$ are expressed in the reference frame of the FC. In \eq{eq:sens-conv-state} and \eq{eq:sens-conv-cov}, the state evolution matrix $\vv{F}_{\tau_m^c - \tau_k^s}$ projects the sensor state/covariance estimates forward by $\tau_m^c - \tau_k^s$, so that they are up to date with the current FC time.
Similarly, the FC also performs a prediction step, for a time duration $T_c$, on all its maintained tracks, by leveraging their motion model. For this, the standard KF prediction equations are used
\begin{equation}\label{eq:pred-central-state}
\vv{C}_{m|m-1}^c(u) = \vv{F}_{T_c} \vv{C}_{m-1|m-1}^c(u) \vv{F}_{T_c}^T + \vv{W},
\end{equation}
\begin{equation}\label{eq:pred-central-cov}
\hat{\vv{x}}_{m|m-1}^c(u) = \vv{F}_{T_c} \hat{\vv{x}}_{m-1|m-1}^c(u).
\end{equation}

\subsubsection{Track association}\label{sec:tt-assoc}
The FC has to compute track-to-track associations before being able to fuse the information from the sensors with its own tracks, as it needs to identify which tracks correspond to the same target. 
There can be \emph{(i)} \emph{sensor-to-center} associations (SC), to verify whether the sensor tracks correspond to any of the maintained central tracks, and \emph{(ii)} \emph{sensor-to-sensor} associations (SS), only for sensor tracks which did not find a SC association, to establish which tracks correspond to the same targets and consequently initialize the correct number of central tracks.
The aim is to find a one-to-one association between two sets of tracks, respectively, indicized by variables $i$ and $j$. Note that $i$ and $j$ may refer to two sensor tracks, in case of an SS association, or to a central track and a sensor track, in case of an SC association.

Initially, SC associations are considered.
As the first step, it is verified whether associations from previous time steps are still valid. To this purpose, unique identifiers associated with each sensor, central track, and sensor track are exploited.
Every sensor-to-center track pair that has a correspondence in previous SC associations is examined to verify whether the association still holds.
This operation consists in computing the Mahalanobis distance \cite{shalom2009probabilistic} between the two tracks and confirming the association only if the distance is lower than, or equal to, a threshold $A_{\rm th}$.
Formally, it is computed as 
\begin{equation}
        \label{eq:mahalanobis}
        \text{d}_M(i,j) = \left(\hat{\vv{x}}(i) - \hat{\vv{x}}(j)\right)^T \vv{P}(i,j)\left(\hat{\vv{x}}(i) - \hat{\vv{x}}(j)\right),
\end{equation}
\begin{equation}
    \label{eq:precision-mat}
    \vv{P}(i, j) = [\vv{C}(i) + \vv{C}(j)]^{-1}
\end{equation}
where $\vv{P}(i, j)$ is the precision matrix inducing the distance.
All track pairs for which $\text{d}_M(i, j) \le A_{\rm th}$ are retained as valid SC associations, while the remaining ones undergo the following further association stages.
Assume $M$ and $N$ central and sensor tracks are left to be associated, respectively.
A $M \times N$ cost matrix, $\vv{\Lambda}$, is obtained, where the value of entry $\Lambda_{ij}$ is computed differently depending on the relationship between central track $i$ and sensor track $j$.
\\
If $i$ and $j$ were previously fused together within a time interval of $T_{\rm th}$ seconds, then, $\Lambda_{ij}$ is computed as in \eq{eq:mahalanobis} with the difference that each track state estimate $\hat{\vv{x}}(q)$ and each covariance matrix $\vv{C}(q)$, $q=i, j$, is replaced by $\hat{\vv{x}}_{\rm dec}(q)=\hat{\vv{x}}(q)-\bar{\vv{x}}(j)$ and $\vv{C}_{\rm dec}(q)=(\vv{C}(q)^{-1}-\bar{\vv{C}}(j)^{-1})^{-1}$, respectively.
$\bar{\vv{x}}(j)$ and $\bar{\vv{C}}(j)$ are the last communicated state and covariance matrix from track $j$, respectively.
$\hat{\vv{x}}_{\rm dec}(q)$ and $\vv{C}_{\rm dec}(q)$ represent the decorrelated versions of the corresponding quantities, according to the \emph{decorrelation principle} \cite{chang1997optimal, chong2000architectures}.
The decorrelation operation removes the effect of previous fusion events that, otherwise, would affect the computation of the association cost \cite{chong2000architectures}.
\\
If $i$ and $j$ were never fused together, or the fusion event happened more than $T_{\rm th}$ seconds before, then, $\Lambda_{ij}=d_M(i, j)$, as in \eq{eq:mahalanobis}, without modifications.
Once matrix $\vv{\Lambda}$ is available, the minimum total cost association is obtained by using, e.g., the Hungarian algorithm \cite{kuhn1955hungarian}, and all associations whose cost doesn't exceed threshold $A_{\rm th}$ are considered as valid SC associations.

After these operations, all acceptable SC associations have been established and only SS associations are left to be computed.
Let $i$ and $j$ be two tracks from different sensors. Then, a similar cost matrix $\vv{\Lambda}$ is built, where $\Lambda_{ij}=d_M(i, j)$, as in \eq{eq:mahalanobis}. Since sensor tracks are originated from different sensors, they are negligibly correlated and there is no need to apply any decorrelation operation on them. Then, the minimum total cost association is obtained, as before, using, e.g., the Hungarian algorithm \cite{kuhn1955hungarian}.
In our experiments we adopted $A_{\rm th}=18$ and $T_{\rm th}=1.3 \times T_c$.

\subsubsection{Radar track fusion algorithm}\label{sec:track-fusion}
The track fusion algorithm behaves differently in case it has to combine two sensor tracks (SS fusion) or one sensor track with a central track (SC fusion). 
If the FC is currently not maintaining any track for a certain subject, but one or more radars are, a new central track needs to be initialized based on the received information from the local sensors. Specifically, two cases may happen: \textit{(i)} if a target $n_1$ is currently tracked by one sensor only, the corresponding central track is initialized using the state and covariance of the sensor track; \textit{(ii)} if the FC receives two tracks that can be associated, say $\mathcal{T}_m^1(n_1)$ and $\mathcal{T}_m^2(n_2)$, these are fused into a single, new track associated with target with central index $u$ (SS fusion). The local tracks from sensors $1$ and $2$ have uncorrelated (or negligibly correlated) errors as they are two sensor tracks, so they can be fused with a weighted combination of their states \cite{chong2000architectures}. The states are weighted by the precision matrices associated with the estimation errors at each sensor. The fusion equations used for the initialization of a new central track, at time $m$, in case \emph{(ii)}, are 
\begin{equation}\label{eq:first-cov}
    \vv{C}_{m|m}^c(u) = \left[\vv{P}_{m}^1(n_1) + \vv{P}_m^2(n_2) \right]^{-1},
\end{equation}
\begin{equation}
\begin{multlined}\label{eq:first-state}
    \hat{\vv{x}}_{m|m}^c(u) = \vv{C}_{m|m}^c(u) \left[\vv{P}_{m}^1(n_1)\hat{\vv{x}}_{m}^1(n_1) + \vv{P}_{m}^2(n_2)\hat{\vv{x}}_{m}^2(n_2) \right],
\end{multlined}
\end{equation}
for the couple of associated tracks $\mathcal{T}^1_m(n_1)$ and $\mathcal{T}^2_m(n_2)$.
Note that, to detect when a SS fusion has to be performed, our algorithm applies the SS association procedure to all the sensor tracks that have \textit{not} been associated to any central track in the current slot.
In case more than $2$ sensors are available, the above process is repeated sequentially using sensors $1$ and $2$ first, then fusing the resulting track with the information from sensor $3$ and so on until the track sets of all $S$ sensors are used.

On the other hand, if the FC has already initialized the track for a subject, the fusion has to be performed between the central track and a sensor track corresponding to the same subject.
Denote by $\boldsymbol{\mathcal{T}}^c_m$ the set of tracks maintained by the FC at the central time step $m$.
Upon receiving the local information $\boldsymbol{\mathcal{T}}^s_m$, the FC runs the track-to-track association algorithm to find pairs of corresponding tracks $\{\mathcal{T}^c_m(u), \mathcal{T}^s_m(n)\}$.
Once such pairs are available, if the timestamp associated with $\boldsymbol{\mathcal{T}}^s_m$ is older than $T_{\rm th}$, then, the same fusion rule of \eq{eq:first-cov} and \eq{eq:first-state} is used, as the two tracks can be considered sufficiently decorrelated, otherwise, each central track is updated with its corresponding sensor track using the \textit{information decorrelation} method \cite{chang1997optimal, chong2000architectures} as follows
\begin{equation}\label{eq:fusion-cov}
    \vv{C}_{m|m}^c(u) = \left[ \vv{P}_{m|m-1}^c(u) + \vv{P}_{m}^s(n) - \bar{\vv{P}}^s(n) \right]^{-1},
\end{equation}
\begin{equation}\label{eq:fusion-state}
\begin{multlined}
    \hat{\vv{x}}_{m|m}^c(u) = \vv{C}_{m|m}^c(u) \left[ \vv{P}_{m|m-1}^c(u) \hat{\vv{x}}_{m|m-1}^c(u)+  \right. \\ \left. + \vv{P}_{m}^s(n)\hat{\vv{x}}_{m}^s(n) - \bar{\vv{P}}^s(n)\bar{\vv{x}}^s(n)\right],
\end{multlined}
\end{equation}
where $\bar{\vv{P}}^s(n)$ and $\bar{\vv{x}}^s(n)$ are the last communicated precision matrix and state estimate of track $\mathcal{T}^s_m(n)$ from sensor $s$ to the FC.
Information decorrelation copes with the problem of time correlated tracks between the FC and the radar sensors, by removing the most recently received information about target $n$ (or $u$ from the FC perspective), as, otherwise, this would be accounted for twice.

\subsubsection{Track initialization and termination}\label{sec:track-man}
To deal with the initialization and termination of central tracks, while keeping the complexity of the system as low as possible, we follow a \mbox{so-called} $\texttt{m}/\texttt{n}$ logic, similar to what is done for the local tracking process of each radar sensor \cite{pegoraro2021real}. Specifically, a track is maintained if it is associated with any of the received sensor tracks for at least $\texttt{m}$ out of the last $\texttt{n}$ frames. Similarly, received sensor tracks that are not associated with any existing central track are initialized as new tracks if they are detected for at least $\texttt{m}$ out of the last $\texttt{n}$ frames. As detailed in \secref{sec:track-fusion}, before initializing a new central track, the received selected tracks form the radars are associated and fused with an SS fusion step, whenever possible. In this way, we avoid multiple initializations of the tracks corresponding to the same targets. 

\section{Experimental results}\label{sec:results}
\begin{figure*}
    \centering
    \subfloat[][in-line\label{fig:setup0-1}]
    {\includegraphics[width=.13\textwidth]{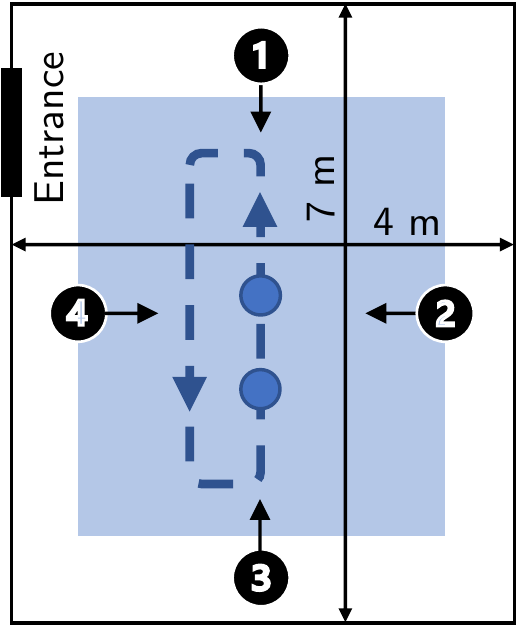}} \,
    \subfloat[][parallel\label{fig:setup0-2}]
    {\includegraphics[width=.13\textwidth]{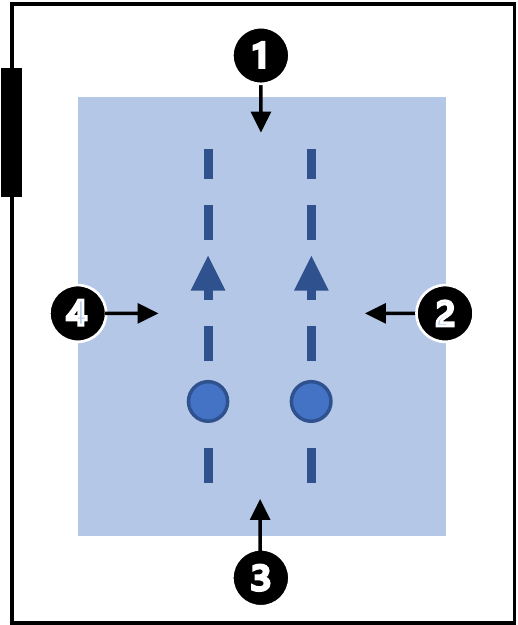}} \,
    \subfloat[][circular\label{fig:setup0-3}]{\includegraphics[width=.13\textwidth]{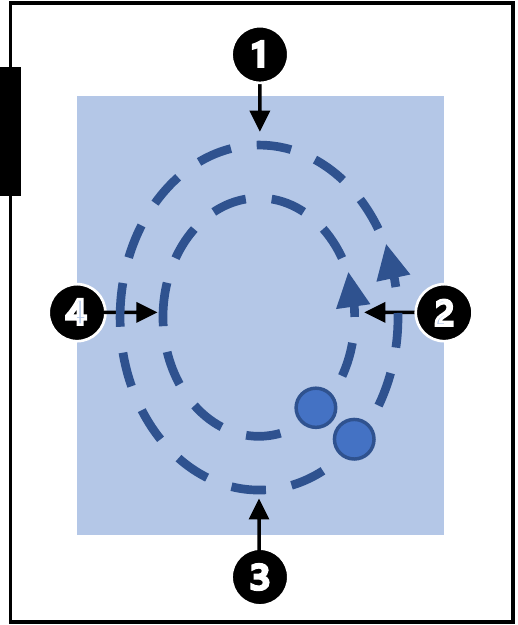}} \,
    \subfloat[][free\label{fig:setup0-4}]{\includegraphics[width=.13\textwidth]{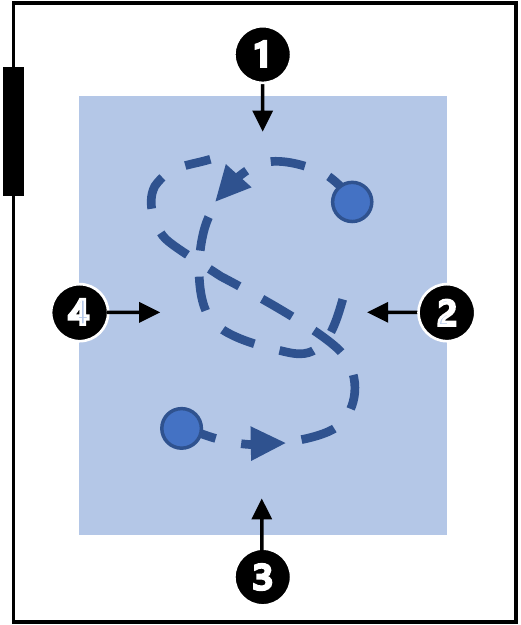}} \,
    \subfloat[][in-line\label{fig:setup0-5}]
    {\includegraphics[width=.13\textwidth]{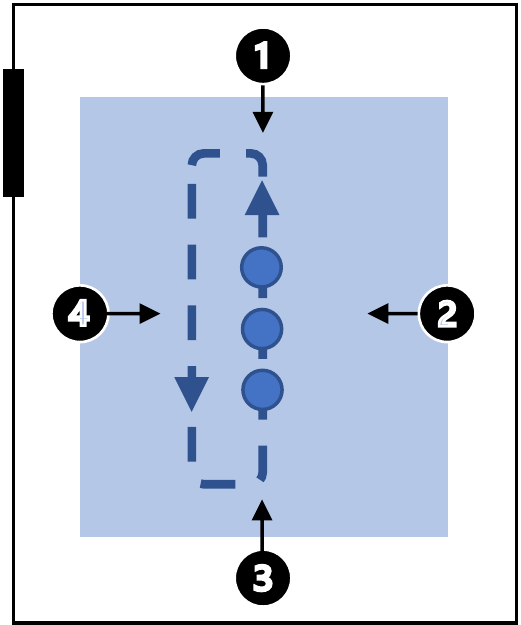}} \,
    \subfloat[][parallel\label{fig:setup0-6}]{\includegraphics[width=.13\textwidth]{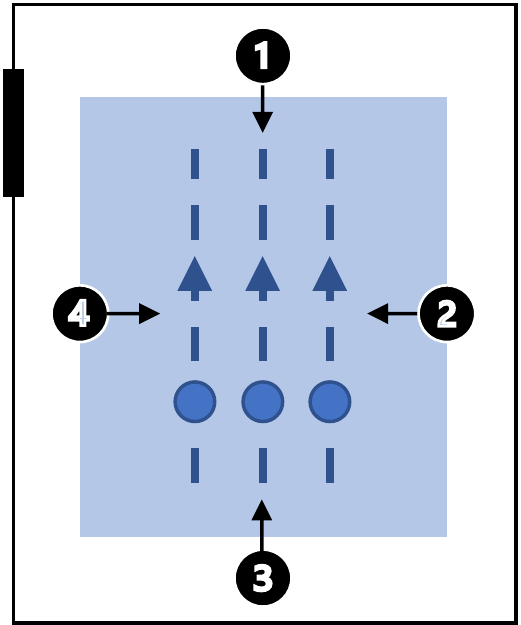}} \,
    \subfloat[][free\label{fig:setup0-7}]{\includegraphics[width=.13\textwidth]{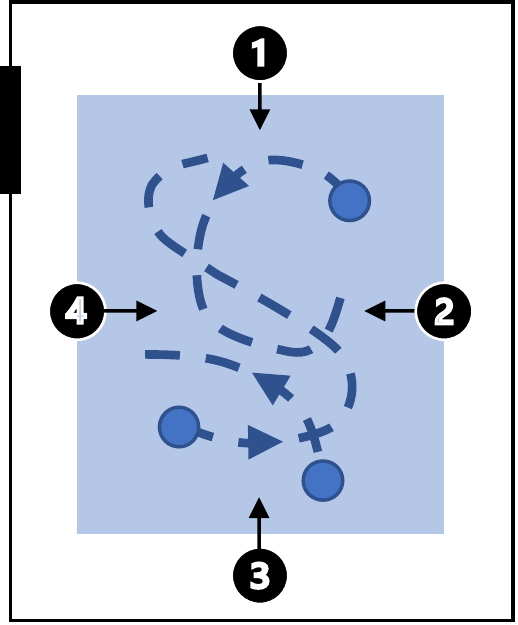}} \\

    \subfloat[][parallel\label{fig:setup1-1}]
    {\includegraphics[width=.13\textwidth]{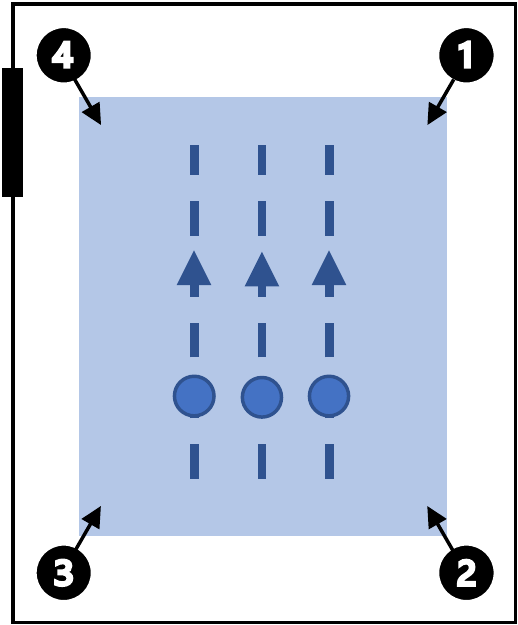}} \,
    \subfloat[][paral-diag\label{fig:setup1-2}]
    {\includegraphics[width=.13\textwidth]{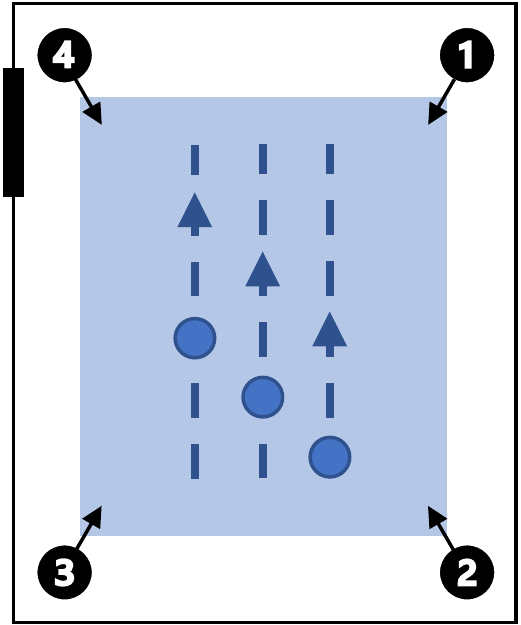}} \,
    \subfloat[][free\label{fig:setup1-3}]
    {\includegraphics[width=.13\textwidth]{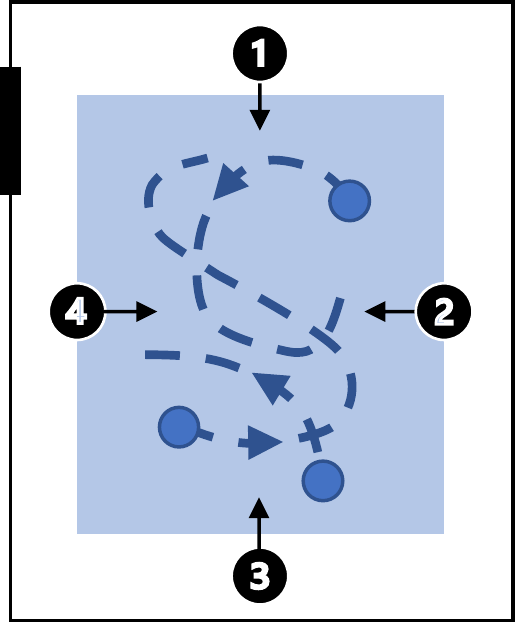}} \,
    \subfloat[][circular\label{fig:setup1-4}]
    {\includegraphics[width=.13\textwidth]{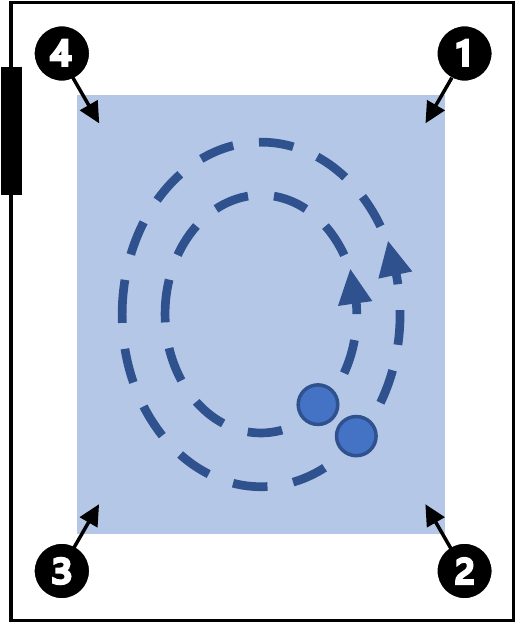}} \,
    \subfloat[][vs-in-line\label{fig:setup1-5}]
    {\includegraphics[width=.13\textwidth]{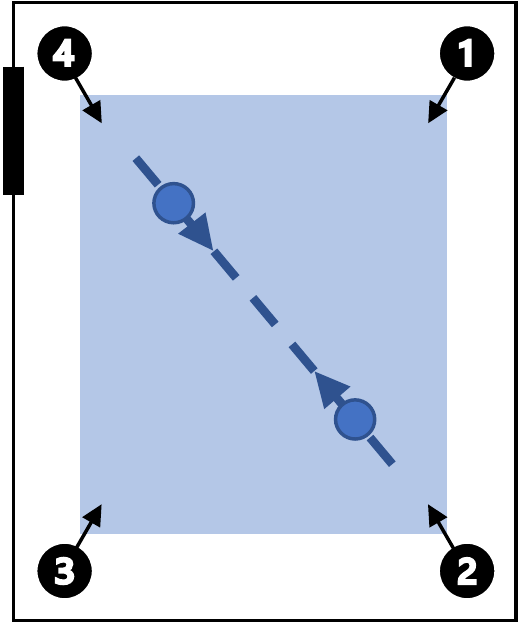}}

     \caption{Setup schemes. The black numbered dots represent the radar devices and the arrows identify their pointing direction. The blue dots represent the moving people, while the dashed lines show the traveled trajectories in the direction given by the blue arrows. The first row shows setup-1 deployments whereas the second row shows setup-2 deployments.}
    \label{fig:setups}
\end{figure*}

\begin{figure}
	\begin{center}   
		\includegraphics[width=8.4cm]{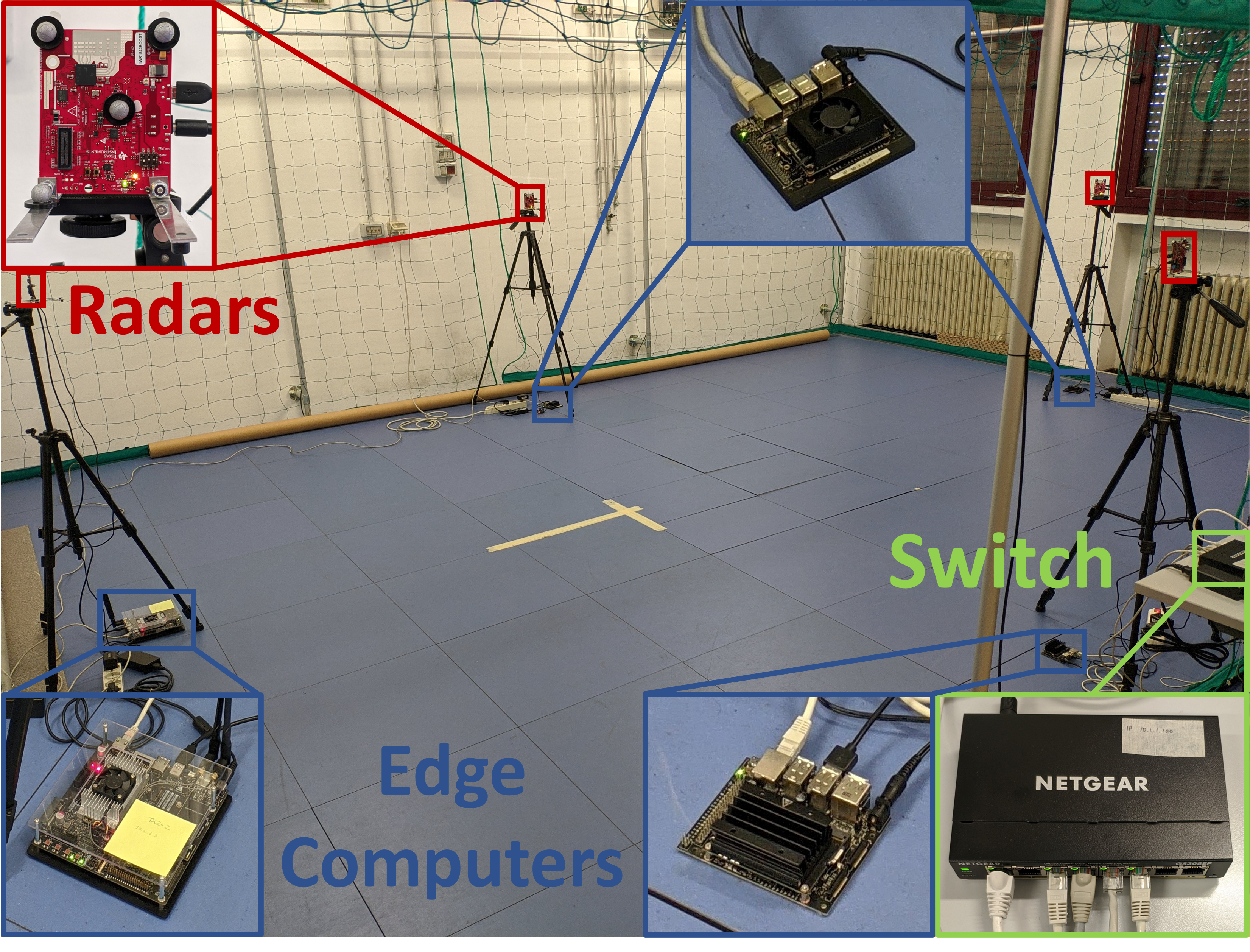} 
		\caption{Experimental setup. The fusion center, not shown in the picture, is connected to the edge computers through a switch.}
		\label{fig:setup-pic}
	\end{center}
\end{figure}

We evaluated ORACLE through an implementation of the RadNet platform \cite{bashirov2022RadNet} featuring 4 Texas Instruments IWR1843BOOST mmWave radars\footnote{https://www.ti.com/tool/IWR1843BOOST} connected to 2 Jetson Xavier NX DevKit\footnote{https://developer.nvidia.com/embedded/jetson-xavier-nx-devkit}, 1 Jetson TX2 DevKit\footnote{https://developer.nvidia.com/embedded/buy/jetson-tx2-developer-kit} and 1 Jetson Nano DevKit\footnote{https://developer.nvidia.com/embedded/buy/jetson-nano-devkit}, all communicating via Ethernet.
The radars operate in the \mbox{$77$-$81$~GHz} band in real-time at a frame rate of \mbox{$1/T_s = 15$~Hz} with a FoV of $\pm 60^{\circ}$ and $\pm 15^{\circ}$ over the azimuth and elevation planes, respectively. \fig{fig:setup-pic} shows a picture of the implemented experimental setup.

\subsection{Implementation notes on numerical stability}
\label{sec:numerical-stability}
In this section, we provide insight into the numerical stability of ORACLE, which are key to implementing the system in practice. 
Specifically, 
during our experiments, we observed that ORACLE's operations on covariance matrices (e.g., the information decorrelation or roto-translations) can easily make them \emph{(i)}~non positive definite and \emph{(ii)}~ill-conditioned (with very large condition numbers), causing wrong track associations and fusion results because of the spoiled inverse matrices.
The solution to point~\emph{(i)} is to enforce positive definiteness by adding a properly designed diagonal matrix.
Formally, let $\vv{P}$  be a symmetric square matrix obtained by ORACLE as part of the information fusion process, and $\lambda_{\min} \le 0$ be its minimum, non-positive eigenvalue.
We obtain the corrected, positive definite matrix $\vv{P}_{\rm pos}$, as
\begin{equation}\label{eq:posdef}
    \vv{P}_{\rm pos} = \vv{P} + (-\lambda_{\min} + \varepsilon)\vv{I}, \quad\varepsilon > 0.
\end{equation}
This approach relies on the fact that a symmetric matrix is positive definite if and only if all its eigenvalues are positive.
As a solution for \emph{(ii)}, we use Ridge regularization \cite{hoerl1970ridge} to limit the condition number of the matrix.
Let $\lambda'_{\min} > 0$ and $\lambda'_{\max} > 0$ be, respectively, the minimum and maximum eigenvalues of the positive definite matrix $\vv{P}_{\rm pos}$. Denote by 
$\mathrm{cond}(\vv{P}_{\rm pos}) = \lambda'_{\max} / \lambda'_{\min}$ the condition number of matrix $\vv{P}_{\rm pos}$, and by $\delta$ the regularization parameter.
The regularized covariance matrix, $\vv{P}_{\rm reg}$, is obtained as
\begin{equation}\label{eq:reg-condnum}
    \vv{P}_{\rm reg} = \frac{1}{1+\delta}(\vv{P}_{\rm pos}+\delta\vv{I}).
\end{equation}
As a result of \eq{eq:reg-condnum}, the minimum and maximum eigenvalues of $\vv{P}_{\rm reg}$ are $\lambda'_{\max}+\delta$ and $\lambda'_{\min}+\delta$, respectively.
To limit the condition number of $\vv{P}_{\rm reg} $, we specify an upper bound for its value, denoted by $c^*$. Then, such bound is enforced by computing  $\delta$ from $\mathrm{cond}(\vv{P}_{\rm reg} )=(\lambda'_{\max}+\delta)/(\lambda'_{\min}+\delta) \le c^*$, which is solved with equality by 
\begin{equation}
    \delta=\max\left(0, \frac{\lambda'_{\max}-c\lambda'_{\min}}{c^*-1}\right).
\end{equation}
In our experiments, we empirically decided to adopt \mbox{$c^*=50$}.  ORACLE applies the correction in \eq{eq:posdef} whenever a covariance (or precision) matrix is non positive definite. The regularization in \eq{eq:reg-condnum}, instead, is used if the condition number of a covariance (or precision) matrix is above~$c^*$.

\subsection{Measurements setup and Dataset}
\label{sec:setup-dataset}
To assess the performance of the proposed method, we conducted tests in a $7\times 4\,\text{m}^2$ research laboratory (see \fig{fig:setup0-1}) equipped with a motion tracking system featuring $10$ cameras. This provides the ground truth (GT) 3D localization of a set of markers placed on 
the radars and on the moving subjects
with millimeter-level accuracy. 
We considered $2$ different scenarios with $4$ radars and $1$, $2$, and $3$ moving targets.
\fig{fig:setups} shows the locations and orientations of the radars in the different setups, where the black numbered dots represent the radar devices and the arrows identify their pointing direction. The blue dots represent the moving people, while the dashed lines show the traveled trajectories in the direction given by the blue arrows. The first row shows setup-1 deployments whereas the second row shows setup-2 deployments.
We also asked the subjects to move according to $6$ possible different trajectories:
\emph{(i)} \emph{in-line}, identifying a movement along a straight line, one subject after the other; \emph{(ii)} \emph{parallel}, identifying a movement along parallel lines; \emph{(iii)} \emph{circular}, corresponding to the two subjects following parallel and circular trajectories; \emph{(iv)} \emph{free}, where all subjects could move freely in the room; \emph{(v)} \emph{paral-diag}, identifying a movement on parallel lines but with the subjects spaced apart along the movement directions; and \emph{(vi)} \emph{vs-in-line}, where the two subjects moved one towards the other following the same linear trajectory. All trajectories are depicted in \fig{fig:setups}. 
In total, we collected $55$ sequences, each $40$~s long.
In every sequence, subjects were tracked by all $4$ radars simultaneously and independently.
Then, tracking information has been fused considering all possible combinations of $1$ to $4$ radars.
However, we experienced the Jetson Nano DevKit edge computer not being able to properly track more than $2$ targets simultaneously, and we noticed the issue after the experiments.
For this reason, in order to provide reliable results, we decided to show results with up to $3$ fused radars.
After filtering out the corrupted data, considering all the evaluated combinations, we analyzed a total of $220$, $187$, and $91$ experiments for $1$, $2$, and $3$ fused radars, respectively.

\subsection{Evaluation metrics}
\label{sec:evaluation-metrics}
To evaluate the self-calibration algorithm performance, we define the \emph{orientation error} as the absolute value of the difference between the true orientation angle of a radar and the estimated one. This is derived from the corresponding rotation matrix, after calibration, as explained in \secref{sec:selfcal-problem}.
The \emph{position error} is defined as the Euclidean distance between the estimated position of the radar and its true position.
In order to assess the tracking performance, we adopt the \emph{Multiple Object Tracking Performance Accuracy} (MOTA) metric, which accounts for the number of misses, false positives, and switches in the object detections, and the \emph{Multiple Object Tracking Performance Precision} (MOTP) metric, which represents the mean position error by considering only correctly tracked objects. More details about these metrics can be found in \cite{bernardin2006multiple}.

\subsection{Self-calibration}
\label{sec:selfcal-res}

As previously mentioned, in the first part of this work we are presenting an enhanced version of mmSCALE \cite{shastri2022mmSCALE}, our self-calibration algorithm. The enhancement, consisting in the addition of the \emph{masking} phase to the self-calibration procedure, allows to handle a wider range of cases, where the previous version was more prone to errors, and to achieve a general improvement in the accuracy of the calibration parameters estimation.
To show the effect of the enhancement, we will focus on the comparison between the old and the new version of the self-calibration algorithm.

\begin{figure}
        \centering
        \includegraphics[width=0.8\columnwidth]{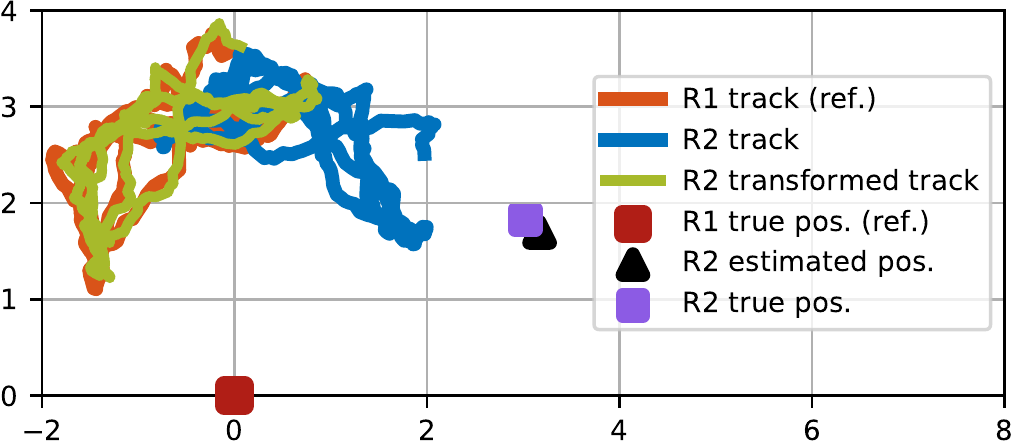}
        \caption{Example of self-calibration with a \emph{free} trajectory.}
        \label{fig:exampleTrack}
\end{figure}

\emph{Qualitative results}.
\fig{fig:exampleTrack} shows a qualitative example of the calibration process.
Here, after finding the optimal rotation and translation parameters, we applied the rigid transformation to the trajectory seen by radar $2$ (blue line, R2), so as to superimpose it with the one of radar $1$ (orange line, R1). The transformed trajectory (green line) matches the reference one well, showing a good calibration result.
We represent the reference radar with a red square (located at $[0, 0]^T$), while the black triangle and the purple square mark the estimated position of radar $2$ and its GT, respectively.
\begin{figure}
        \centering
        \includegraphics[width=1\columnwidth]{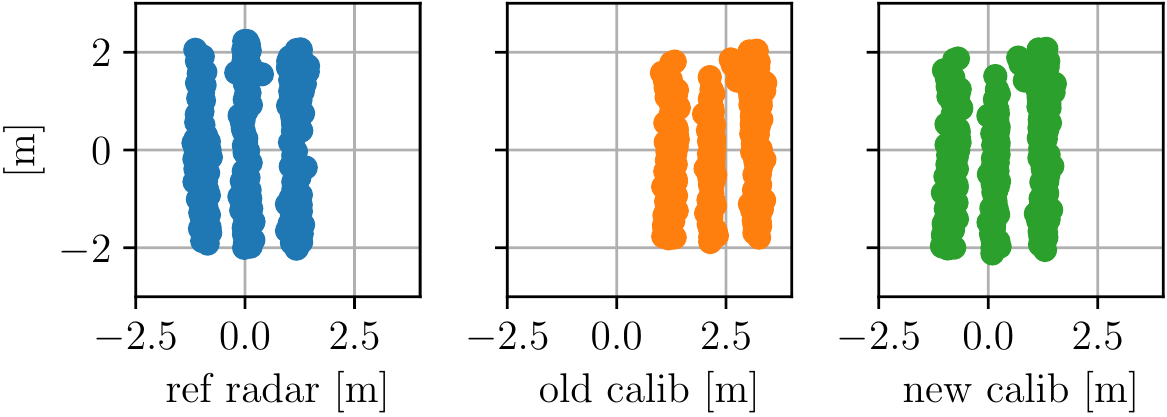}
        \caption{Example of a situation where the old algorithm fails while the new one doesn't. In this case, linear, similar trajectories led to a mistake in the track association phase, resulting in a shift in the estimated location of the radar.}
        \label{fig:self-example}
\end{figure}

As long as only one target is being tracked, the track association phase is easy and it is likely that no errors occur. If the number of tracked targets increases, the track association phase becomes more challenging. Our track association cost (see \secref{sec:tracks-association}) is able to lead to a correct association for most situations. However, there are some particular cases where it is not sufficient. \fig{fig:self-example} shows an example of one of such cases with $3$ subjects following a \emph{parallel} trajectory.
The blue lines represent the trajectories of the reference radar in its RS, the orange lines shows the trajectories of another radar after transforming them in the reference radar's RF using the old self-calibration algorithm, while the green lines represent the trajectories from the same radar after transforming them using the new self-calibration algorithm.
In this scenario, all trajectories are very similar, as the subjects proceed in parallel and at the same speed. Because of the very high correlation between the track positions over time, the association costs are very similar and the final association result depends on subtle numerical variations due to the tiny differences between the trajectory shapes. From the figure, we notice that this causes a wrong association between the tracks from the reference radar and the other one, reflecting in a shifted estimate of the radar's RS, when using the old method. The new method, instead, correctly copes with this situation.

\begin{figure}
        \centering
        \includegraphics[width=1\columnwidth]{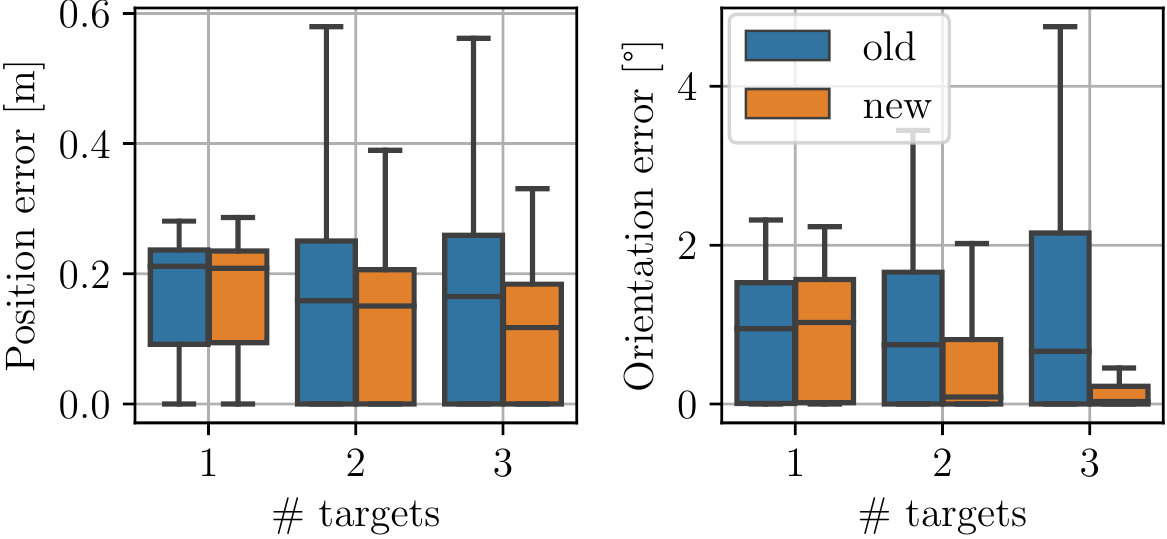}
        \caption{Comparison of the self-calibration results when using the old and the new self-calibration algorithm as a function of the number of targets tracked during the calibration phase.}
        \label{fig:self-results}
\end{figure}

\emph{Position and orientation errors}. \fig{fig:self-results} shows a comparison between the calibration performance using the old and the new version of the self-calibration algorithm versus the number of targets tracked. When only one target is tracked, there is almost no difference between the two algorithms, while a clear improvement is observed as the number of targets increases. In particular, the new algorithm has a great effect in reducing the sparsity of the box plots, meaning it is increasing the number of cases it is able to handle correctly. \tab{tab:self-comparison} shows the numerical results in terms of median and interquartile range (IQR). It also shows the difference between the new and the old algorithm.

\subsection{Fusion center tracking accuracy} \label{sec:fc-acc}
\begin{figure*}
        \centering
        \includegraphics[width=1\textwidth]{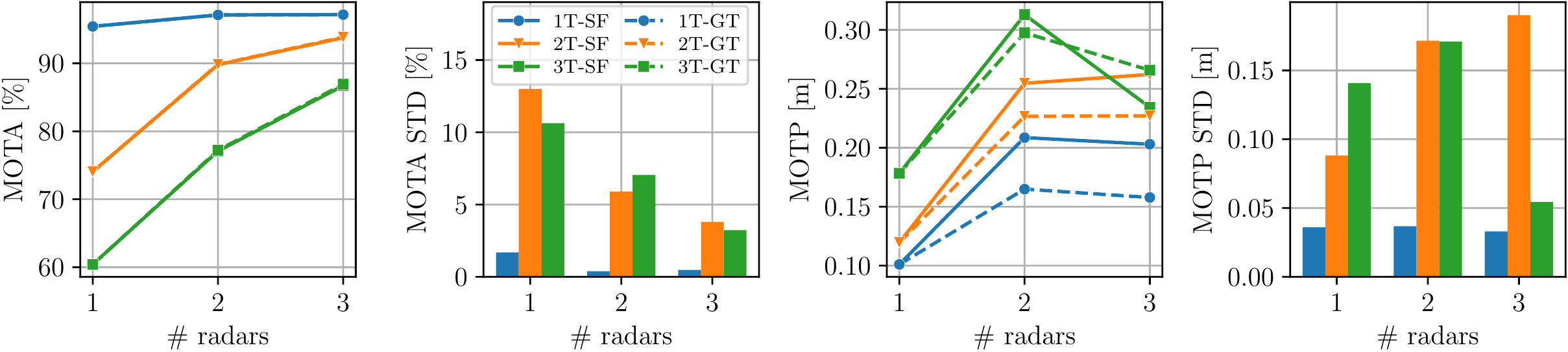}
        \caption{Average MOTA and MOTP as a function of the number of radars used for the fusion and of the number of targets tracked. Solid and dashed lines identify, respectively, results obtained using self-calibration (SF) and ground-truth (GT) to estimate radars' location and orientation.
        The bar charts represent the 1 standard deviations for the SF case.}
        \label{fig:tracking-results}
\end{figure*}

We evaluate the performance of the fusion algorithm in two cases: \emph{(i)} using the roto-translation parameters obtained from the GT; and \emph{(ii)} using their estimations from the self-calibration algorithm. In order to evaluate ORACLE in a more realistic scenario, when using self-calibration, the transformation parameters are computed only once per setup-trajectory pair,
that is, the first time a sequence with that setup-trajectory pair is elaborated. Then, all sequences with the same setup and trajectory type use the same parameters.
\fig{fig:tracking-results} shows the tracking results versus the number of targets tracked and the number or radars used for the fusion. In the figure, xT-SF and xT-GT denote a case where x targets are tracked and, respectively, self-calibration or GT are used.
The bar charts represent the 1 standard deviations for the self-calibration results.
The values are computed as the average over all the sequences with the same number of radars and targets.
Numerical results using self-calibration are presented in \tab{tab:motmetrics}.

\begin{table}
\centering
\caption{Comparison of self-calibration algorithms}
\label{tab:self-comparison}
\scalebox{.87}{
    \begin{tabular}{ccccccc}
        \toprule        
        \multicolumn{2}{c}{\textbf{Median Error (IQR)}} & \textbf{Old} & \textbf{New} & \textbf{New-Old Diff.} \\
        \midrule
         \multirow{2}{*}{\textbf{1T}}   & \textbf{Position [m]}           & $0.21 (0.14)$ & $0.21 (0.14)$ & $\pm0.00(\pm0.00)$\\
                                        & \textbf{Orientation [$^\circ$]} & $0.95 (1.50)$ & $1.00 (1.60)$ & $+0.05(-0.10)$\\
        \midrule
         \multirow{2}{*}{\textbf{2T}}   & \textbf{Position [m]}           & $0.16 (0.25)$ & $0.15 (0.21)$ & $-0.01(-0.04)$\\
                                        & \textbf{Orientation [$^\circ$]} & $0.75 (1.70)$ & $0.09 (0.81)$ & $-0.66(-0.89)$\\
        \midrule
        \multirow{2}{*}{\textbf{3T}}    & \textbf{Position [m]}           & $0.17 (0.26)$ & $0.12 (0.18)$ & $-0.05(-0.08)$\\
                                        & \textbf{Orientation [$^\circ$]} & $0.66 (2.20)$ & $0.03 (0.22)$ & $-0.63(-1.98)$\\
        \bottomrule
    \end{tabular}}
\end{table}
\begin{table}
\centering
\caption{Summary of ORACLE's tracking performance}
\label{tab:motmetrics}
\scalebox{0.87}{
    \begin{tabular}{cccccc}
        \toprule
        & & \textbf{1R} & \textbf{2R} & \textbf{3R} & \textbf{3R-1R diff.} \\
        \midrule
         \multirow{2}{*}{\textbf{1T}} & \textbf{MOTA [\%]} & $95 \pm 2$ & $97 \pm 0$ & $97 \pm 0$ & $+2$ \\
                                  & \textbf{MOTP [m]} & $0.10 \pm 0.04$ & $0.21 \pm 0.04$ & $0.20 \pm 0.03$ & $+0.10$ \\
        \midrule
         \multirow{2}{*}{\textbf{2T}} & \textbf{MOTA [\%]} & $74 \pm 13$ & $90 \pm 6$ & $94 \pm 4$ & $+20$ \\
                                  & \textbf{MOTP [m]} & $0.12 \pm 0.09$ & $0.25 \pm 0.17$ & $0.26 \pm 0.19$ & $+0.14$ \\
        \midrule
         \multirow{2}{*}{\textbf{3T}} & \textbf{MOTA [\%]} & $60 \pm 11$ & $77 \pm 7$ & $87 \pm 3$ & $+27$ \\
                                  & \textbf{MOTP [m]} & $0.18 \pm 0.14$ & $0.31 \pm 0.17$ & $0.23 \pm 0.05$ & $+0.05$ \\
        \bottomrule
    \end{tabular}
    }
\end{table}

When only one target is tracked, there is almost no difference between using a single radar or multiple fused radars ($+2\%$).
As the number of tracked people increases, single sensors experience a remarkable decrease in the MOTA ($-35\%$) while the FC maintains high performance, with a MOTA as high as $87\%$ when $3$ targets are tracked, leading to an improvement with respect to single sensors of $+27\%$.
This is due to the fact that multiple targets may often create occlusions with respect to single radars, increasing the number of misses and switches in the tracks.
Instead, occlusions can be mitigated by fusing data from different points of view. 
We also note that GT and self-calibration achieve very similar results in terms of MOTA.

In general, MOTP slightly increases in the fused tracking.
The reason is twofold. First, noise can be incorporated during the fusion process and slightly affect the localization performance. Second, MOTP is computed only for correctly tracked targets. Because a single radar's tracking capability is limited, successful tracking occurs only in sufficiently simple scenarios, where MOTP would be straightfowardly low. On the contrary, multiple radars track targets successfully even in more complicated cases. This increases the range of points for which we compute MOTP to include more arduous and inevitably less precise location and movement estimates.
Interestingly, for the $1$T and $2$T cases, after increasing from $1$ to $2$ radars, MOTP is almost constant when moving from $2$ to $3$ fused radars, suggesting that this could be the case also if more radars are fused.
Following the $3$T lines, instead, MOTP increases from $1$ to $2$ radars and then decreases when $3$ radars are used, reaching the same values of the $2$T lines. A possible explanation for this is that $2$ radars are not enough for tracking $3$ targets, leading to errors in the track associations that cause the MOTP to increase. $3$ radars, instead, have better tracking capabilities and can better handle $3$ targets.
For the most challenging scenario ($3$ targets), MOTP is $31$~cm when $2$ radars are used and $23$~cm when $3$ radars are used. 
MOTP is slightly lower when using GT rather than self-calibration because of the more precise knowledge about sensors' position and orientation.

As a final test, we acquired some sequences where we fuse all of the $4$ radars for various $2$-target trajectories, providing a MOTA and MOTP of $95\%$ and $20$~cm, respectively, while single radars on the same sequences reach a MOTA and MOTP of $78\%$ and $11$~cm, respectively. Since, for this test, we only collected a few sequences that do not represent a statistically significant set, we present them only as an example.

These results show that our self-calibration algorithm works well in combination with the proposed fusion algorithm and that they can be effectively used together to enable an occlusion-resilient people tracking through a self-calibrated radar network, requiring almost no human intervention.

\subsection{Robustness to reduced fusion rate} \label{sec:frate-res}

\begin{figure}
    \centering
    \includegraphics[width=1\columnwidth]{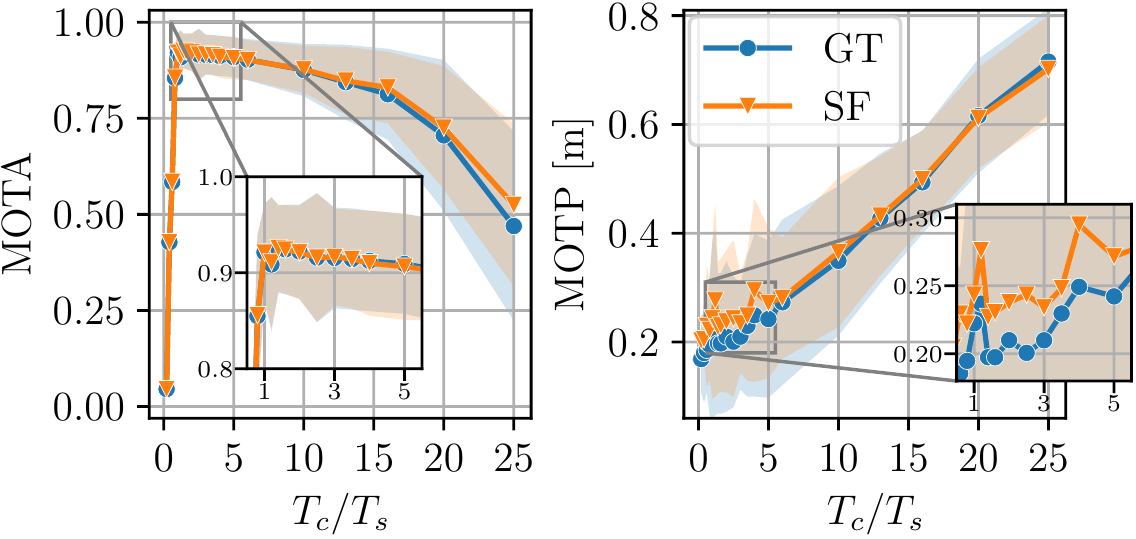}    
    \caption{Average MOTA and MOTP versus different values of the ratio $T_c/T_s$ when calibration is performed using GT measurements (GT), or with the self-calibration (SF). The shaded areas represent $1$ standard deviation.     \label{fig:change-slot-time}}
\end{figure}

In certain resource-constrained applications it may be useful to reduce the FC processing rate of the sensors data, in order to lower the computational burden. However, this requires striking a balance between fusion rate and tracking accuracy, as decreasing the processing rate reduces the capability of the FC to follow the movement of the subjects.\\
In \fig{fig:change-slot-time}, we show the MOTA and MOTP curves as a function of the ratio $T_c/T_s$. This is varied by fixing $T_s = 66.7$~ms ($15$~fps) and changing $T_c$ from $0.2T_s$ to $25T_s$. Values are obtained by averaging all experiments with $3$ radars. We can identify three regions.\\
\noindent \textit{(i)} For $T_c/T_s < 0.8$, the MOTA is very low, as the FC runs significantly faster than the sensors and, therefore, has to rely mostly on the KF predictions, which are inaccurate after a few consecutive steps. The MOTP instead is unaffected as it is obtained only on the successfully tracked subjects.\\
\noindent \textit{(ii)} For $0.8 \leq T_c/T_s \leq 5$, our system achieves the best performance in terms of MOTA, i.e., over $90\%$. At the same time, the MOTP is still low, with errors of less than $29.4$~cm when using self-calibration with $T_c/T_s = 5$. This shows that, if necessary, the processing load on the FC can be \textit{reduced by $5$ times} with negligible performance degradation.\\
\noindent \textit{(iii)} For $T_c/T_s > 5$, MOTA degrades slowly and MOTP increases. This is because the time-step of the FC, especially towards the end of this region, is too long to accurately follow human movement using the CV model.  \\
Finally, we notice that the MOTA is almost unaffected by using self-calibration in place of the GT sensor locations and orientations. This holds for all regions \textit{(i)-(iii)}. However, as expected, the MOTP is slightly worse in case self-calibration is used, as the residual error in the locations of the sensors indirectly affects the FC tracking precision.

\subsection{Effect of radars' location on fused tracking}
\begin{figure}
        \centering
        \includegraphics[width=1\columnwidth]{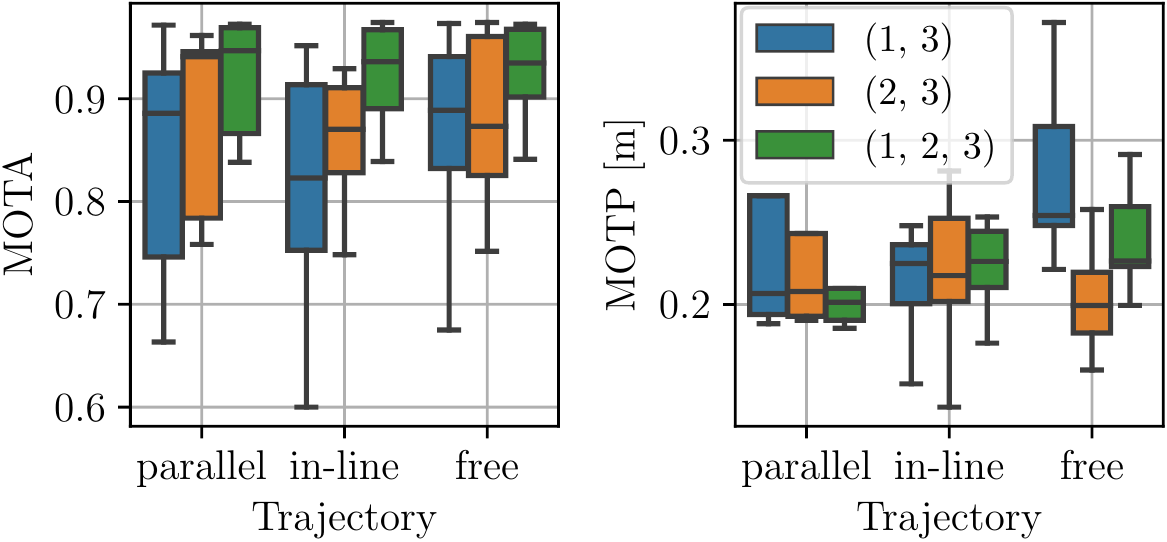}
        \caption{Box plots of MOTA and MOTP for particular trajectories, as a function of specific combinations of fused radar. More precisely, $(1,3)$ has facing radars, $(2, 3)$ has perpendicular radars, ($1, 2, 3$) fuses all of the three. Radar numbers correspond to those in \fig{fig:setups}, first row (setup-1).}
        \label{fig:traj-comparison}
\end{figure}
When tackling the problem of people tracking through multiple radars, it is interesting to explore how different radar deployments affect the results. This kind of study is beyond the scope of this paper and would require a denser deployment of radars. For this reason, here we show only some preliminary results, while we leave a deeper inspection of the problem to future developments. In \fig{fig:traj-comparison} we compare the results for $3$ trajectories and some specific combinations of radars for the fusion. In particular, according to setup-1 (see \fig{fig:setups}, first row) combination $(1, 3)$ corresponds to two radars facing each other, combination $(2, 3)$ has perpendicular radars, while combination $(1, 2, 3)$ fuses all of the three. 
For each combination, results are computed using self-calibration and averaging over all sequences featuring the particular trajectory, with either $1$, $2$, or $3$ targets.
Considering MOTA, perpendicular radars are generally better than facing radars, while fusing $3$ radars always provides the best results.
Regarding MOTP, there is no clear trend common to all trajectories. The median MOTP is always within the $[0.20, 0.25]$~m interval, which means there are no great differences depending on the combinations. The only case worth mentioning is that featuring free trajectory and $(1, 3)$ combination, where MOTP values are generally higher than in the other cases.
In conclusion, from this brief analysis, it appears that \emph{(i)} a larger number of radars is to be preferred over a lower one, and \emph{(ii)} radars with more diverse points of view provide, in general, better tracking results.

\section{Conclusions} \label{sec:concl}
In this work, we presented ORACLE, a solution to the mmWave radar network deployment and integration problem for human sensing purposes.
First, ORACLE automatically estimates the relative position and orientation of the radars with respect to a common reference system. Then, it exploits such estimates to fuse the information about people tracked by different radars at a fusion center, enhancing the resilience of the subject localization in case of occlusions.
ORACLE estimates the radars' position and orientation with a median error of $0.12$~m and $0.03^{\circ}$, respectively, exploiting the movement trajectories of tracked people. With respect to existing self-calibration techniques, ORACLE is more robust to multiple subjects concurrently moving in the environment, with no need to follow any predetermined trajectory for the calibration.
By fusing multiple radars tracking information, ORACLE improves on single sensors by up to $27\%$ in mean tracking accuracy, with a mean precision of $23$~cm in the most challenging case of $3$ targets moving.
Finally, ORACLE handles different time steps for the single sensors and for the FC, keeping the tracking accuracy higher than $90\%$ when the ratio between the central and the sensors time step is $0.8 \le T_c/T_s \leq 5$.
These results substantiate ORACLE as a key technology enabler for distributed people tracking with radar networks, serving as a base system for a large variety of applications, from personnel recognition to restricted areas monitoring, elderly care, customer profiling, and many others.

Future research includes the extension of ORACLE to multiple disjoint radar networks (where none of the radars of a network share a part of the FoV with any of the radars of the other networks), and the study of how different points of view of the same scene affect the tracking performance.

\section*{Acknowledgment}
This work received support from the European Commission's Horizon 2020 Framework Programme under the Marie Sk{\l}odowska-Curie Action MINTS (GA no.~861222).\\
The authors are grateful to Laura Canil for designing and creating the graphical abstract. 

\bibliographystyle{IEEEtran}
\bibliography{IEEEabrv,biblio}

\end{document}